\documentclass[twopage,11pt] {article} 
\usepackage{graphicx}
\usepackage{bm}

\usepackage{amssymb,multicol}
\usepackage{amsmath}

\begin{document}

\title{Black Holes with Skyrme Hair}
\author{Noriko Shiiki and Nobuyuki Sawado \\
{\it Department of Physics, Tokyo University of Science, Noda, Chiba 
278-8510, Japan }}
\pagestyle{myheadings} 
\thispagestyle{plain}         
\markboth{Your Name}{Noriko Shiiki and Nobuyuki Sawado} 
\setcounter{page}{1}

\maketitle

\begin{abstract}
This paper is intended to give a review of the recent 
developments on black holes with Skyrme hair. 

The Skyrme model is an effective meson theory where the baryons 
are identified with topological solitons, so-called skyrmions. 
The baryon number $B$ corresponds to the topological charge. 
Thus the model gives a unified description of hadron physics 
in terms of meson fields. 
The spherically symmetric $B=1$ skyrmion and axially symmetric 
$B=2$ skyrmion were found numerically.    
Upon collective quantization, they produce correct nucleon 
and deuteron observables respectively. 
Approximate solutions with higher baryon numbers were also constructed by using 
the rational map ansatz. 
Depending on the baryon number, they exhibit various discrete symmetries with 
striking similarities to BPS monopoles. 

The Einstein-Skyrme system is also known to possess black hole solutions 
with Skyrme hair. The spherically symmetric black hole skyrmion with $B=1$ was 
the first discovered counter example of the no-hair conjecture for black holes.  
Recently we found the $B=2$ axially symmetric black hole skyrmion.  
In this system, the black hole at the center of the skyrmion 
absorbs the baryon number partially, leaving fractional charge outside the horizon. 
Therefore the baryon number is no longer conserved.  
We examine the $B=1, 2$ black hole solutions in detail in this paper. 
 
The model has a natural extension to the gauged version which can 
describe monopole black hole skyrmions.   
Callan and Witten discussed the monopole catalysis of proton decay 
within the Skyrme model. 
We apply the idea to the Einstein-Maxwell-Skyrme system and 
obtain monopole black hole skyrmions.  
Remarkably there exist multi-black hole skyrmion solutions 
in which the gravitational, electromagnetic, and strong forces 
between the monopoles are all in balance.   
The solutions turn out to be stable under spherically symmetric linear 
perturbations. 
\end{abstract}
\pagebreak

\section{Introduction}
Application of the Skyrme model to the physics of the early Universe or
equivalent high-energy physics is an interesting subject. 

The attempts to explain (anti-)baryon production in high-energy collisions   
within the Skyrme model were made in Refs.~\cite{degrand84,ellis88,ellis89}. 
The scenario for the production is the same as the one for producing other 
topological defects such as monopoles, cosmic strings or domain walls in 
the early Universe via the Kibble mechanism~\cite{kibble76,turok89}. 
Those defects are formed after a phase transition in which 
global symmetry is broken spontaneously. 
In the case of skyrmions, broken is the chiral 
symmetry and as a result one of the possible orientations of 
the chiral field is chosen randomly in the internal symmetry space 
to form the defects. The use of the Skyrme model is expected to 
describe nonperturbative phenomena to which perturbative QCD is not 
accessible. 

Another interesting application of the Skyrme model is the monopole 
catalysis of proton decay. 
The monopole catalysis of proton decay was studied firstly at the lepton 
and quark level by Callan~\cite{callan82}, Rubakov~\cite{rubakov82} 
and Wilczek~\cite{wilczek82} independently. 
In the grand unified theory there was a symmetry between baryons and leptons 
at energy scale $10^{15}$ GeV and therefore it would have been possible 
for the following reaction to be driven through the anomaly  
\begin{eqnarray}
	 p+{\rm Monopole} \rightarrow e^{+}+{\rm pions}+{\rm Monopole}\,. \label{}
\end{eqnarray}
Interestingly this reaction occurs at a strong interaction rate unsuppressed 
by any small-coupling-constant effects. 
Callan and Witten proposed not quark-monopole but proton-monopole 
interactions by using the Skyrme model~\cite{callan-witten84} to make a more 
realistic estimation of the catalysis cross section possible.  
The decay process can be explained because 
in the presence of a monopole the baryon number is no longer equal 
to the topological charge of the meson field. In fact, there exist 
non-topological solitons with non-zero baryon number which can decay 
without topological problems. 
The extension to non-abelian monopole catalysis was also studied in 
Ref.~\cite{brihaye01}. 

For the study of the high-energy physics in the early Universe, 
it may be important to consider the effects of gravity on baryons. 
With the present value of the gravitational constant, those effects are 
insignificant upto the Planck energy $10^{19}$ GeV. 
In fact, in the Einstein-Skyrme theory, the Planck mass is related to 
the pion decay constant $F_{\pi}$ and the coupling constant $\alpha$ 
by $M_{pl}=F_{\pi}\sqrt{4\pi/\alpha}$. To realize the realistic value 
of the Planck mass, the coupling constant should be extremely small with 
$\alpha \sim O(10^{-39})$, which makes the theory little different from 
the theory without gravity.     
However, some theories such as scalar-tensor gravity theory  
and Kaluza-Klein theory predict the time variation 
of the gravitational constant~\cite{brans61,marciano84,barrow02}. 
And also theories with extra dimensions predict that a true Planck scale 
is of order a TeV. Thus there may have been an epoch in the early Universe 
where the gravitational effects on baryons were significant. We consider those 
effects worth being studied in the Skyrme model. 
The advantage of the Skyrme model to more realistic nucleon models 
is that it is straightforward to couple to gravity. 
Although investigating the Einstein-Skyrme system is not expected to 
provide any quantitatively reliable results, we cannot exclude the 
possibility that it may give some qualitative insight into the 
gravitational interaction of baryons. 

The Einstein-Skyrme system has been studied by various authors. 
The first obtained solutions in this system are spherically symmetric 
black holes with Skyrme hair~\cite{luckock86,luckock87,droz91}. 
Later, regular solutions for $B=1$~\cite{luckock87,droz91,bizon92} and axially 
symmetric black hole and regular solutions for $B=2$~\cite{shiiki02} were found. 
The extended models to $SU(3)$ and $SU(N)$ were also studied in Refs.~\cite{zakrzewski04}.  
It has been observed that microscopic black holes can support 
Skyrme hair admitting fractional baryon number outside the horizon. 
This configuration can be interpreted as a skyrmion partially 
absorbed by the microscopic black hole. 
Therefore the model provides a semiclassical framework to study 
the absorption rate of a proton by a black hole of comparable size. 
However this process is rather insignificant because black holes 
of the size of a proton have large fluxes of Hawking radiation~\cite{hawking75}. 
Situations in which baryon decay process might become more realistic and 
significant over the Hawking radiation occurs when the black hole carries 
electric or/and magnetic charge with which the skyrmion interacts electromagnetically 
as well as gravitationally. Since a charged black hole in general has an effective 
temperature lower than that of a Schwarzschild black hole of the same mass, 
the Hawking radiation effect is less. Especially interesting is the extremal 
black hole which has a vanishing effective temperature, so the Hawking radiation 
may even vanish~\cite{gibbons75}.  
According to this speculation, we analyzed monopole black hole solutions 
with Skyrme hair~\cite{moss-shiiki00}. This model provides the semiclassical 
framework in which to study monopole black hole catalysis of proton decay. 
Although macroscopic charged black holes perhaps do not exist in nature, 
microscopic charged black holes may have been created in the early Universe 
and remain as stable relics today. Indeed the GUT monopole has a typical 
radius $R_{M}\sim M_{X}^{-1}$ and a mass $g^{-2}M_{X}$, the Schwarzschild 
radius for the mass is then $R_{BH}\sim 2Gg^{-2}M_{X}$ and thus  
$R_{M}/R_{BH}\sim 2Gg^{-2}M_{X}^{2}$ where $g^{2}$ is the coupling 
constant renormalized at the $X$-boson mass scale $M_{X}$. 
Hence for a sufficiently large magnetic charge $p=1/g$ and 
heavy gauge boson mass, the monopoles could have undergone gravitational 
collapse to form monopole black holes. This monopole black hole has an 
additional internal degree of freedom for electric charges being a dyon 
black hole analogous to the 't Hooft-Polyakov monopoles being the 
Julia-Zee dyon~\cite{julia-zee75}. 
    
In subsequent sections we review the recent developments on black 
holes with Skyrme hair.    
Section $2$ contains an introduction of the Skyrme model along  
with its brief historical review. 
In section $3$ we review the $B=1$ black hole skyrmion solution and 
its stability analysis.  
In section $4$ the Einstein-Skyrme model coupled to abelian 
gauge fields is studied to obtain $B=1$ monopole black hole skyrmion 
solutions. Its stability is examined in detail and the technique to 
study monopole black hole catalysis of baryon decay is discussed briefly.  
$B=2$ black hole skyrmions are examined in section $5$. It is shown that 
the energy density and baryon density of the solutions are torus in shape 
having a horizon at the center. 
Conclusion and Discussion are in section $6$. 
    
Throughout this paper we use the metric signature $(-,+,+,\cdots)$ 
and the Einstein summation convention. Greek indices are used 
to denote spacetime components of a tensor, while Italian indices 
are used to denote purely spatial components. 

\pagebreak 

\section{The Skyrme Model and Skyrmions}
In this section we give an introduction of the Skyrme model,  
reviewing its historical development. 
A more detailed and complete review of the Skyrme model 
is provided by Zahed and Brown in Ref.~\cite{brown86}.  

\subsection{Historical Review}
It has been known that in the large-$N_{c}$ effective meson theory, baryons emerge 
as solitons. This recognition has a long history. 

In the Standard Model QCD is the theory for strong interactions. However it is 
$SU(3)$ gauge theory and there is no hope of solving it exactly and therefore 
an approximation method is necessary for the qualitative estimation of QCD. 
The most convenient approximation method would be a perturbation which, in fact, 
worked very well for QED. However it seemed that there was no small parameter 
to expand in low-energy QCD since the coupling constant is no longer small 
at low energies. 
The unexpected possible parameter was proposed by 't Hooft~\cite{thooft74}. 
Although in the real world hadrons have definite color degrees of freedom 
$N_{c}=3$, he suggested considering the number of colors $N_{c}$ as a free 
parameter and obtaining the series expansion in powers of $1/N_{c}$. 
Then a detailed analysis of the diagrams shows that the limit of 
$N_{c}\rightarrow\infty$ fixing $g_{3}^{2}N_{c}$ is equivalent to the effective 
chiral field theory for mesons where $g_{3}$ is a coupling constant. 
Thus 't Hooft succeeded to prove that the chiral effective Lagrangian 
is indeed low energy QCD. The next question was, how can baryons 
be incorporated in this large $N_{c}$ effective meson theory?  

Witten gave a rigorous answer to this question~\cite{witten79}. 
In his paper he showed that baryon mass is of order $N_{c}$ which is 
inverse of the expansion parameter $1/N_{c}$. 
Hence in the large-$N_{c}$ limit, they become singular. This is a typical 
feature seen in solitons. From this fact, Witten conjectured that 
baryons are solitons in the large-$N_{c}$ meson theory. 

Much earlier than Witten, in the flow of development of the effective 
theory of QCD, Skyrme attempted to construct a unified theory of 
strong interactions based on the meson field alone~\cite{skyrme58}. 
He explicitly constructed the minimal Lagrangian if the non-linear 
sigma model in $d=1+3$ that supports topological soliton solutions 
and suggested that they are baryons~\cite{skyrme61}. 
However this remarkable idea was neglected until the $80$s. 

A unification of the ideas came in 1983. Witten studied solitons 
in the current algebra effective Lagrangian, which is the $SU(N_{c})$ 
chiral models with the QCD anomalous term (Wess-Zumino 
term~\cite{wess71})~\cite{witten83-2}. In QCD, the decay process 
$K{\bar K}\rightarrow \pi^{+} +\pi^{-}+\pi^{0}$ is anomalous 
in the sense that it is observed in nature but is forbidden 
in the chiral model because of the symmetry $U\rightarrow U^{\dagger}$ 
which conserves the parity of the number of mesons. 
By including the Wess-Zumino term in the action, the anomalous process 
can be correctly described~\cite{witten83}. 
Witten showed that the anomalous piece of the current is responsible 
for the soliton solutions and their quantum numbers such as spin 
and baryon number. In particular for the $N_{c}=2$ case which is the chiral 
model with $N_{c}$ flavor degrees of freedom, the Wess-Zumino term vanishes. 
However from the homotopy argument $\pi_{3}(SU(2))=Z$, there are still 
solitons. Besides $\pi_{4}(SU(2))=Z_{2}$ suggests that in a suitably 
compactified space-time, there are two topological classes of maps 
from space-time to $SU(2)$ and hence the solitons can be spin-$\frac{1}{2}$ 
particles~\cite{finkelstein68}. A detailed analysis for the property 
of the skyrmion as a nucleon was performed in Ref.~\cite{adkins83} 
upon quantization of its collective coordinates. 

\subsection{The Model}

In the lowest order, the chiral Lagrangian is given by 
\begin{eqnarray}
	{\cal L}=\frac{F_{\pi}^{2}}{16}\,{\rm tr}\,(U^{\dagger}\partial_{\mu}U
	U^{\dagger}\partial_{\mu}U) \;,\;\;\; U=e^{2i{\vec \tau}\cdot
	{\vec \pi}/F_{\pi}} \label{}
\end{eqnarray}
in the exponential parameterization where $F_{\pi}=186$ MeV is the pion 
decay constant, ${\vec \tau}$ are the Pauli matrices and ${\vec \pi}$ 
are pion fields. 
The static energy is then given by 
\begin{eqnarray}
	E=\int d^{3}x \,\frac{F_{\pi}^{2}}{16}\,{\rm tr}\,(U^{\dagger}\partial_{i}U
	U^{\dagger}\partial_{i}U)\,. \label{}
\end{eqnarray}
One can show that this model does not support a topological soliton 
as follows. Let us introduce the dimensionless variable ${\tilde {\bf x}}$ 
defined by ${\bf x}=\alpha {\tilde {\bf x}}$. Then for a static configuration, 
the energy can be written as 
\begin{eqnarray}
	E=\alpha \int d^{3}{\tilde x}\, \frac{F_{\pi}^{2}}{16}\,{\rm tr}\,
	(U^{\dagger}\partial_{i}UU^{\dagger}\partial_{i}U)\,. \label{}
\end{eqnarray}
The integrand on the right-hand side is non-negative and hence the 
energy is minimized at $\alpha =0$ with $E=0$. Therefore the solution 
is trivial. 

Skyrme introduced a new term in the fourth order derivative which 
retains the chiral symmetry of the model but supports a soliton 
solution~\cite{skyrme61}. The so-called Skyrme model is defined by 
\begin{eqnarray}
	{\cal L}_{S}=\frac{F_{\pi}^{2}}{16}\,{\rm tr}\,(U^{\dagger}\partial_{\mu}U
	U^{\dagger}\partial_{\mu}U)+\frac{1}{32a^{2}}\,{\rm tr}\,[\partial_{\mu}U
	U^{\dagger},\partial_{\nu}UU^{\dagger}]^{2} \label{}
\end{eqnarray}
where $a$ is a dimensionless parameter whose value can be fixed by 
experiment. The static energy is then given by 
\begin{eqnarray}
	E= \int d^{3}x \left\{\frac{F_{\pi}^{2}}{16}\,{\rm tr}\,(U^{\dagger}
	\partial_{i}UU^{\dagger}\partial_{i}U)
	+\frac{1}{32a^{2}}\,{\rm tr}\,[\partial_{i}U
	U^{\dagger},\partial_{j}UU^{\dagger}]^{2}\right\} \,.  \label{sta-ene}
\end{eqnarray}
One can see that this model indeed supports a nontrivial soliton solution 
by following the earlier argument. Expressing the static energy in terms 
of the rescaled variable ${\tilde {\bf x}}$, one obtains 
\begin{eqnarray}
	E= \alpha \int d^{3}{\tilde x} \,\frac{F_{\pi}^{2}}{16}\,{\rm tr}\,
	(U^{\dagger}\partial_{i}UU^{\dagger}\partial_{i}U)
	+\frac{1}{\alpha}\int d^{3}{\tilde x}\, 
	\frac{1}{32a^{2}}\,{\rm tr}\,[\partial_{i}UU^{\dagger},\partial_{j}
	UU^{\dagger}]^{2} \,.  \label{}
\end{eqnarray}
Since both integrands on the right-hand side are nonnegative, the energy 
is minimized at $\alpha \neq 0$. 

Let us find a topological soliton in the Skyrme model. We introduce 
$A_{\mu}=U^{\dagger}\partial_{\mu}U$ and write the energy in terms of $A_{\mu}$ 
\begin{eqnarray}
	E=\int d^{3}x\,{\rm tr}\,\left\{\frac{F_{\pi}^{2}}{16}
	A_{i}A_{i}^{\dagger}+\frac{1}{32a^{2}}(\epsilon_{ijk}A_{j}A_{k})
	(\epsilon_{ilm}A_{l}A_{m})^{\dagger}\right\}\,. \label{energy-a}
\end{eqnarray}
The boundary condition for $E$ to be finite is 
\begin{eqnarray}
	A_{i}\rightarrow 0 \;\;\;\;\; {\rm as}\; |{\bf x}|\rightarrow \infty \label{}
\end{eqnarray}
which is equivalent to saying that $U$ approaches some constant matrix at infinity. 
Without loss of generality, we define this constant as the unit matrix 
\begin{eqnarray}
	U\rightarrow I \;\;\;\;\; {\rm as} \; |{\bf x}|\rightarrow \infty\,. \label{}
\end{eqnarray}
Then the spacetime is compactified to the three-sphere $S^{3}$. 

There is a lower bound for the energy~(\ref{energy-a}). From the 
Cauchy-Schwarz inequality,  
\begin{eqnarray}
	\left(\frac{F_{\pi}}{4}A_{i}-\frac{1}{4a}\epsilon_{ijk}
	A_{j}A_{k}\right)^{2}\ge 0\, , \label{}
\end{eqnarray}
one can obtain 
\begin{eqnarray}
	E\ge \frac{F_{\pi}}{8a}\int d^{3}x\,|{\rm tr}\,(\epsilon_{ijk}
	A_{i}A_{j}A_{k})|\,. \label{ege}
\end{eqnarray}
The topological current of the skyrmion is 
\begin{eqnarray}
	B^{\mu}=-\frac{\epsilon^{\mu\nu\rho\sigma}}{24\pi^{2}}
	\,{\rm tr}\,(A_{\nu}A_{\rho}A_{\sigma})  \label{}
\end{eqnarray}
with $\partial_{\mu}B^{\mu}=0$. The topological charge is 
given by the zeroth component of the current 
\begin{eqnarray}
	B=\int d^{3}x\, B^{0}=-\frac{1}{24\pi^{2}}\int d^{3}x\,
	{\rm tr}\,(\epsilon_{ijk}A_{i}A_{j}A_{k}) \label{b1}
\end{eqnarray}
which corresponds to the winding number of the map $S^{3}\rightarrow S^{3}$ 
and characterized by an integer as $\pi_{3}(S^{3})=Z$. 
In the Skyrme model, topological charge is identified with 
the baryon number and hence skyrmions are interpreted as baryons~\cite{skyrme61}. 

From Eqs.~(\ref{ege}) and (\ref{b1}) one finds the Bogomol'nyi bound 
\begin{eqnarray}
	E \ge \frac{3\pi^{2}F_{\pi}}{a}|B|\,. \label{}
\end{eqnarray}
At present, no soliton solution saturating this bound is found. 
But the skyrmions which have been obtained numerically probably 
represent the global minimum of the energy for given $B$. 

\begin{figure}
\hspace{2cm}
\includegraphics[height=6.5cm, width=8.5cm]{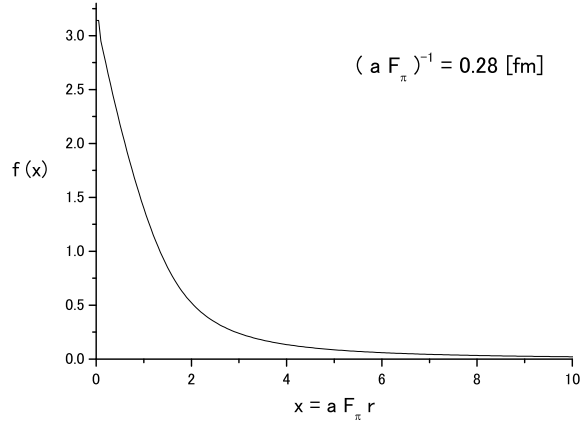}
\caption{\label{fig-profile} Profile $f$ as a function of 
          radial distance $x=aF_{\pi}r$.}
\end{figure} 

The Skyrmion can be found by using the hedgehog ansatz 
\begin{eqnarray}
	U=e^{if(r){\hat {\bf x}}\cdot\tau}=\cos f(r)+i{\vec n}\cdot 
	 {\vec \tau} \sin f(r) \label{}
\end{eqnarray}
where ${\vec n}={\bf x}/r$ with the boundary conditions 
\begin{eqnarray}
	f(0) = \pi \;, \;\;\; f(\infty)=0 \,.\label{bou-con}
\end{eqnarray}
Inserting the ansatz into the energy functional~(\ref{sta-ene}), 
one gets 
\begin{eqnarray}
	E=4\pi \int_{0}^{\infty} dr r^{2}\left[\frac{F_{\pi}^{2}}{8}
	\left(f'^{2}+\frac{2\sin^{2}f}{r^{2}}\right)
	+\frac{1}{2a^{2}}\frac{\sin^{2}f}{r^{2}}\left(\frac{\sin^{2}f}{r^{2}}
	+2f'^{2}\right)\right] \label{}
\end{eqnarray}
where a prime denotes derivative with respect to $r$. The static solution 
can be given as an extremum of the energy. Hence it satisfies $\delta E/\delta 
F =0$. Equivalently
\begin{eqnarray}
 	\left(\frac{x^{2}}{4}+2\sin^{2}f\right)f''
	+\frac{x}{2}f'+f'^{2}\sin 2f
	-\frac{\sin^{2}f\sin 2f}{x^{2}}=0 \label{skyrme-eq}
\end{eqnarray}
where we introduced a dimensionless variable $x=aF_{\pi}r$. 
The solution of the equation (\ref{skyrme-eq}) satisfying the boundary 
conditions (\ref{bou-con}) is shown in Fig.~\ref{fig-profile}.  
For this solution, we have 
\begin{eqnarray}
	B=-\frac{1}{2\pi}[2f-\sin 2f]_{\pi}^{0}=1\,. \label{}
\end{eqnarray}
Thus this solution represents a nucleon. 
For higher baryon numbers, 
$B=2$ skyrmions were obtained numerically and shown to be axially 
symmetric~\cite{kopelio-87,braaten88}. 
Braaten {\it et al.} constructed skyrmions upto $B=6$ by 
descretizing the model on a cubic lattice~\cite{braaten-90}. 
Interestingly, it has been shown that multi-skyrmions with $B > 2$ exhibit 
various discrete symmetries analogously to multi-BPS 
monopoles in the use the rational map ansatz proposed 
in Ref.~\cite{houghton98}.
 
\pagebreak 

\section{$B=1$ Black Hole Skyrmions} 

As stated earlier, it is straightforward to 
couple the Skyrme model to gravity. 
The Skyrme Lagrangian and topological current are written 
in a covariant manner. 
After imposing the suitable ansatz and boundary 
conditions on both the chiral field and metric, 
one can solve numerically the Einstein equations coupled 
to the chiral field to obtain black hole solutions with Skyrme hair.  
The Einstein-Skyrme system was firstly studied by Luckock and 
Moss~\cite{luckock86} where the Schwarzschild black hole with 
Skyrme hair was obtained numerically. 
This is a counter example of the no-hair conjecture for black 
holes~\cite{ruffini71}.  
They observed that the presence of the horizon in the core of skyrmion 
unwinds the skyrmion, leaving fractional baryonic charge outside the horizon. 
The full Einstein-Skyrme system was solved later to obtain spherically 
symmetric black holes with Skyrme hair~\cite{luckock87,droz91} and regular 
gravitating skyrmions~\cite{droz91,bizon92}. 
In this section we review the $B=1$ black hole skyrmion solution  
and analyze its stability following Refs.~\cite{luckock86,luckock87,
droz91,bizon92,heusler92}.  

\subsection{Field Equations and Static Solutions}
The Einstein-Skyrme system is defined by the Lagrangian
\begin{eqnarray}
	{\cal L} &=& {\cal L}_{G}+{\cal L}_{S} \label{action} \\
	&=& \frac{R}{16\pi G} +\frac{F_{\pi}^{2}}{16}g^{\mu\nu}{\rm tr}
	\,(L_{\mu}L_{\nu})+\frac{1}{32a^{2}}g^{\mu\nu}g^{\rho\sigma}{\rm tr}\,
	([L_{\mu},L_{\rho}][L_{\nu},L_{\sigma}])\nonumber 
\end{eqnarray} 
For $B=1$, let us impose the hedgehog ansatz on the chiral field
\begin{eqnarray}
	U(r)=\cos f(r)+i{\vec n}\cdot{\vec \tau}\sin f(r) \,. \label{}
\end{eqnarray}
Correspondingly, we shall impose the spherically symmetric ansatz on 
the metric 
\begin{eqnarray}
	ds^{2}=-N^{2}(r)C(r)\, dt^{2}+\frac{1}{C(r)}dr^{2}+r^{2}d\Omega^{2} \label{metric}
\end{eqnarray}
where we have defined 
\begin{eqnarray*}
	C(r)=1-\frac{2Gm(r)}{r} \, .
\end{eqnarray*}
At the horizon $r=r_{h}$, we have $C(r_{h})=0$, that is, $m(r_{h})=r_{h}/(2G)$. 
Inserting these ansatz into the Lagrangian (\ref{action}), one obtains the static 
energy density for the chiral field 
\begin{eqnarray}
	E_{S}=\frac{4\pi F_{\pi}}{a} \int_{x_{h}}^{\infty} 
	\left\{\frac{1}{8}\left(Cf'^{2}
	+\frac{2\sin^{2}f}{x^{2}}\right)+\frac{\sin^{2}f}{2x^{2}}
	\left(2Cf'^{2}+\frac{\sin^{2}f}{x^{2}}\right)\right\}N x^{2}\, dx \, . \label{energy}
\end{eqnarray}
where we have introduced dimensionless variables   
\begin{eqnarray*}
	x=aF_{\pi}r =r/0.28 \,{\rm fm} \; , \;\;\;\;\; {\hat m} (x)=aF_{\pi}Gm(r)\, . 
\end{eqnarray*}
It should be noted that in the presence of a black hole, the matter field 
is defined only outside the horizon and therefore the integral over the 
space is performed from the horizon to infinity. 

The covariant topological current is defined by  
\begin{eqnarray}
	B^{\mu}=-\frac{\epsilon^{\mu\nu\rho\sigma}}{24\pi^{2}}\frac{1}{\sqrt{-g}}
	{\rm tr}\left(U^{-1}\partial_{\nu}UU^{-1}\partial_{\rho}UU^{-1}
	\partial_{\sigma}U\right). 
	\label{topological_current} \label{baryon_current}
\end{eqnarray}
whose zeroth component corresponds to the baryon number density 
\begin{eqnarray}
	B^{0}=-\frac{1}{2\pi^{2}}\frac{1}{N}\frac{f'\sin^{2}f}{r^{2}}\, . \label{}
\end{eqnarray}
We impose the boundary conditions on the profile function
\begin{eqnarray}
	f(\infty)=0 \, ,  \label{}
\end{eqnarray} 
which determines the value at the horizon $f(r_{h})=f_{h}$.  
Then the baryon number becomes    
\begin{eqnarray}
	B=\int \sqrt{-g}\,B^{0} \, d^{3}x = -\frac{2}{\pi}\int_{f_{h}}^{0}
	\sin^{2}f df = \frac{1}{2\pi}(2f_{h}-\sin 2f_{h}) \, . \label{}
\end{eqnarray}
This shows that the solution possesses fractional baryonic charge 
when $f_{h} < \pi$. 

\begin{figure}
\hspace{2cm}
\includegraphics[height=7.5cm, width=9.5cm]{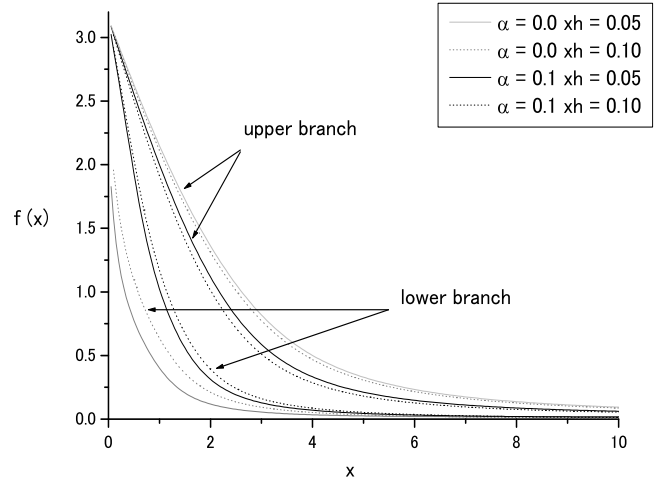}
\caption{\label{fig:f} Profile function $f (x)$ with $\alpha =0.00, 0.10$ 
and $x_{h}=0.05, 0.10$ for upper and lower branch.}
\end{figure}

\begin{figure}
\hspace{2cm}
\includegraphics[height=7.5cm, width=9.5cm]{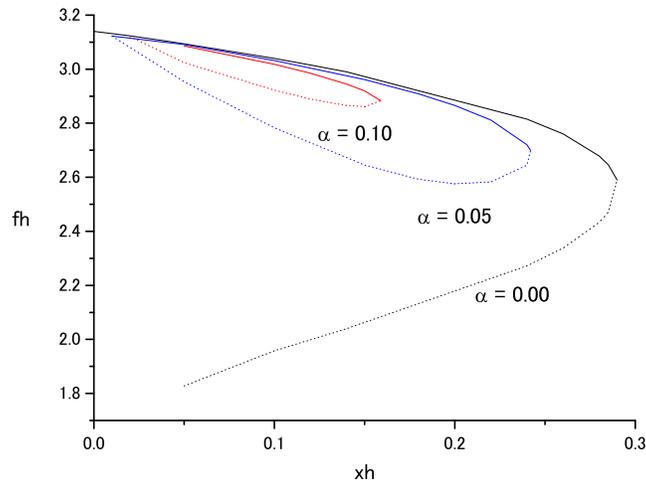}
\caption{\label{fig:xh-fh} The shooting parameter $f_{h}$ 
as a function of the horizon size $x_{h}$ with $\alpha =0.00, 0.05, 0.10$. 
Solid line shows the stable branch and dotted line the unstable branch.  }
\end{figure}

The field equations for the gravitational fields $N(x)$ and ${\hat m}(x)$ 
can be derived from the Einstein equations 
\begin{eqnarray}
	G_{\mu\nu}=8\pi G T_{\mu\nu} \label{}
\end{eqnarray}
which read 
\begin{eqnarray}
	N'&=& \frac{\alpha}{4}\left(x+\frac{8\sin^{2}F}{x}\right)Nf'^{2} \label{n-eq} \\
	{\hat m}'&=& \frac{\alpha}{8} \left[(x^{2}+8\sin^{2}f)Cf'^{2}
	+2\sin^{2}f+\frac{4\sin^{4}f}{x^{2}}\right] \label{m-eq}
\end{eqnarray} 
where we have defined the coupling constant $\alpha = 4\pi GF_{\pi}^{2}$. 
The variation of the static energy (\ref{energy}) with respect to the 
profile $f(x)$ leads to the field equation for matter 
\begin{eqnarray}
	&& f''= \frac{1}{NC(x^{2}+8\sin^{2}f)}\left[-(x^{2}+8\sin^{2}f)N'Cf'
	+\left(1+\frac{4\sin^{2}f}{x^{2}}+4Cf'^{2}\right) \right. \nonumber \\
	&& \left. \times N\sin 2f -2(x+4\sin 2ff')NCf'
	-2\left(1+\frac{8\sin^{2}f}{x^{2}}\right)
	({\hat m}-{\hat m}'x)Nf' \right] \, .\label{f-eq}
\end{eqnarray}
To solve these coupled field equations, let us consider the boundary 
conditions for the gravitational fields. Expanding the fields 
$f(x),{\hat m}(x),N(x)$ around the horizon $x=x_{h}$ and substituting 
into the field equations~(\ref{n-eq})-(\ref{f-eq}), one obtains upto 
second order 
\begin{eqnarray*}
	f(x)&=&f_{h}+f_{1}(x-x_{h})+O((x-x_{h})^{2}) \\
	{\hat m}(x)&=&\frac{x_{h}}{2}+{\hat m}_{1}(x-x_{h})+O((x-x_{h})^{2}) \\
      N(x)&=& N_{h}+N_{1}(x-x_{h})+O((x-x_{h})^{2}) \end{eqnarray*}
where $f_{h}$ and $N_{h}$ are shooting parameters which should be determined 
so as to satisfy the boundary conditions at infinity $f(\infty)=0$ and $N(\infty)=1$, 
and 
\begin{eqnarray}
	{\hat m}_{1}&=&\frac{\alpha}{4}\left(\sin^{2}f_{h}
	+\frac{2\sin^{4}f_{h}}{x_{h}^{2}}\right) \\
	f_{1}&=&\frac{(x_{h}^{2}+4\sin^{2}f_{h})\sin 2f_{h}}{x_{h}(x_{h}^{2}
	+8\sin^{2}f_{h})(1-2{\hat m}_{1})} \\
	N_{1}&=&\frac{\alpha}{4}\left(x_{h}+\frac{8\sin^{2}f_{h}}{x_{h}}\right)
	N_{h}f_{1}^{2} \, .\label{}
\end{eqnarray}
The dependence of the profile function on the coupling constant and 
the horizon size is shown in Fig.~\ref{fig:f}. 
There are two branches of solutions for each value of the coupling constant. 
Let us call the solution with larger (smaller) values of $f_{h}$ as upper 
(lower)-branch.  
The skyrmion shrinks as $\alpha$ or/and $x_{h}$ increase for 
upper-branch while it expands for lower-branch. 
Fig.~\ref{fig:xh-fh} shows the dependence of the value of 
the profile function at the horizon on the horizon size. 
One can see that for $\alpha\neq 0$, all black hole 
solutions converge to globally regular solutions as 
$x_{h}\rightarrow 0$. 
As $\alpha$ approaches to zero, the lower-branch solutions converge 
to the Schwartzschild black hole with Skyrme hair. 
On the other hand, the upper-branch solutions converge to the 
$n=1$ colored black hole solution~\cite{bizon90}. 
The maximum value of the coupling constant with which black hole 
solutions exit is $0.126$. 

\pagebreak 

\subsection{Linear Stability Analysis}
The linear stability of the $B=1$ black hole skyrmion 
was studied in detail in Refs.~\cite{heusler92,bizon92}. 
In order to examine the stability, let us consider 
the time-dependent Skyrme action given by 
\begin{eqnarray}
	{\cal S}=-\frac{\pi e^{2}F_{\pi}^{4}}{2}\int \left[(-\frac{1}{e^{\delta}C}
	{\dot f}^{2}+Cf'^{2})u + v \right]e^{\delta} dx \label{time-action}
\end{eqnarray}
where we have defined 
\begin{eqnarray}
	\delta={\rm log}N \;, \;\;\; 
	u=x^{2}+8\sin^{2}f \; , \;\;\; v=\sin^{2}f(x^{2}+2\sin^{2}f) \label{}
\end{eqnarray}
Then time-dependent field equation can be obtained by taking variation 
with respect to $f$ as  
\begin{eqnarray}
	(e^{\delta}Cuf')'+\frac{1}{2}\left(\frac{1}{e^{\delta}C}{\dot f}^{2}
	-e^{\delta}Cf'^{2}\right)u_{f}-\frac{e^{\delta}v_{f}}{x^{2}}
	=\frac{1}{e^{\delta}C}u{\ddot f} \label{time-field-eq}
\end{eqnarray}
where $u_{f}=\delta u/\delta f$ and $v_{f}=\delta v/\delta f$. 
From the action (\ref{time-action}), the time-dependent 
Einstein equations are derived as 
\begin{eqnarray}
	G_{00}=8\pi GT_{00} & \rightarrow &
	1-C-C'x=\frac{\alpha^{2}}{4}\left[\left(\frac{1}{e^{2\delta}C}{\dot f}^{2}
	+Cf'^{2}\right)u+\frac{2v}{x^{2}}\right]\\
	G_{11}=8\pi GT_{11} & \rightarrow &
	-1+C+\frac{(e^{2\delta}C)'}{e^{2\delta}}x=\frac{\alpha^{2}}{4}
	\left[\left(\frac{1}{e^{2\delta}C}{\dot f}^{2}+Cf'^{2}\right)u
	-\frac{2v}{x^{2}}\right] \label{}
\end{eqnarray}
which read to the following two equations for the gravitational fields 
\begin{eqnarray}
	&& \delta'= \frac{\alpha^{2}}{4x}\left(\frac{1}{e^{2\delta}C^{2}}{\dot f}^{2}
	+f'^{2}\right)u \label{stab_d} \\ 
	&& -(Cx)'+1 = \frac{\alpha^{2}}{2x^{2}}v+C\delta'x \label{stab_c} \,.\label{}
\end{eqnarray}
Let us consider the small radial fluctuations of the profile and gravitational 
fields around the static classical solutions denoted by $f_{0}$, 
$N_{0}$ and $C_{0}$ as
\begin{eqnarray}
	f(x,t)&=& f_{0}(x)+f_{1}(x,t) \\
	\delta(x,t)&=& \delta_{0}(x)+\delta_{1}(x,t) \\
	C(x,t)&=& C_{0}(x)+C_{1}(x,t) \, .\label{}
\end{eqnarray}
Substituting into Eqs.~(\ref{stab_d}) and (\ref{stab_c}) gives the linearized equations   
\begin{eqnarray}
       && \delta_{1}' = \frac{\alpha^{2}}{2x}(2u_{0}f_{0}'f_{1}'
       +u_{f_{0}}f_{0}'^{2}f_{1}) \label{ddel1} \\
	 && -(e^{\delta_{0}}C_{1}x)'=\frac{\alpha^{2}}{2x^{2}}e^{\delta_{0}}
	v_{f_{0}}f_{1}+e^{\delta_{0}}C_{0}\delta_{1}'x \, .\label{dc1}
\end{eqnarray}
Eq.~(\ref{ddel1}) and the classical field equation derived 
from Eq.~(\ref{time-field-eq})
\begin{eqnarray}
	 \frac{e^{\delta_{0}}v_{f_{0}}}{x^{2}}=(e^{\delta_{0}}C_{0}u_{0}f_{0}')'
	-\frac{1}{2}e^{\delta_{0}}C_{0}u_{f_{0}}f_{0}'^{2} \label{static-field}
\end{eqnarray}
are inserted into Eq.~(\ref{dc1}) and resultantly one gets  
\begin{eqnarray}
	-(e^{\delta_{0}}C_{1}x)'=\frac{\alpha^{2}}{2}(e^{\delta_{0}}
	C_{0}u_{0}f_{0}'f_{1})'  \label{}
\end{eqnarray}
which can be integrated immediately  
\begin{eqnarray}
	C_{1}=-\frac{\alpha^{2}}{2x}C_{0}u_{0}f_{0}'f_{1} \,. \label{c1}
\end{eqnarray}
Similarly we shall linearize the field equation~(\ref{time-field-eq}). 
Using Eqs.~(\ref{ddel1}), (\ref{static-field}) and (\ref{c1}), one arrives at   
\begin{eqnarray}
	(e^{\delta_{0}}C_{0}u_{0}f_{1}')'-U_{0}f_{1}=\frac{1}{e^{\delta_{0}}C_{0}}
	{\ddot f_{1}} \label{f1-eq}
\end{eqnarray}
where 
\begin{eqnarray}
	U_{0}&=&-(e^{\delta_{0}}C_{0}u_{f_{0}}f_{0}')'
	+\left(\frac{\alpha^{2}}{2x}e^{\delta_{0}}C_{0}u_{0}^{2}
	f_{0}'^{2}\right)'-\frac{\alpha^{2}}{2x}e^{\delta_{0}}C_{0}
	u_{0}u_{f_{0}}f_{0}'^{3} \nonumber \\
	&& +\frac{1}{2}e^{\delta_{0}}C_{0}u_{ff_{0}}f_{0}'^{2}
	+\frac{e^{\delta_{0}}v_{ff_{0}}}{x^{2}}\,. \label{}
\end{eqnarray}
Setting $f_{1}=\xi (x)e^{i\omega t}/\sqrt{u_{0}}$ , Eq.~(\ref{f1-eq}) becomes 
\begin{eqnarray}
	-(e^{\delta_{0}}C_{0}\xi')'+\left[\frac{1}{2\sqrt{u_{0}}}
	\left(e^{\delta_{0}}C_{0}\frac{u_{0}'}{\sqrt{u_{0}}}\right)'
	+\frac{1}{u_{0}}U_{0}\right] \xi =\omega^{2}
	\frac{1}{e^{\delta_{0}}C_{0}}\xi \,. \label{eigen-eq}
\end{eqnarray}
Let us introduce the tortoise coordinate $x^{*}$ such that 
\begin{eqnarray}
	\frac{dx^{*}}{dx}=\frac{1}{e^{\delta_{0}}C_{0}} \label{}
\end{eqnarray}
with $-\infty < x^{*} < +\infty$. Eq.~(\ref{eigen-eq}) is then reduced 
to the Strum-Liouville equation 
\begin{eqnarray}
	-\frac{d^{2}\xi}{dx^{*2}}+{\hat U_{0}}\xi =\omega^{2}\xi  \label{}
\end{eqnarray}
where
\begin{eqnarray}
	{\hat U_{0}}=e^{\delta_{0}}C_{0}\left[\frac{1}{2\sqrt{u_{0}}}
	\left(e^{\delta_{0}}C_{0}\frac{u_{0}'}{\sqrt{u_{0}}}\right)'
	+\frac{1}{u_{0}}U_{0}\right] \,. \label{strum}
\end{eqnarray}
If the black hole skyrmion is stable, this equation has no negative mode 
which induces exponential grow in $\xi$.  
Numerically solving the Eq.~(\ref{eigen-eq}) or (\ref{strum}), 
one finds no negative mode in the upper branch and one negative 
mode in the lower branch.  
Therefore, it is concluded that the black hole skyrmion in the upper 
branch is stable and unstable in the lower branch. 

\pagebreak

\section{Monopole Black Hole Skyrmions}
In this section, charged black hole solutions with Skyrme hair 
are discussed. We find stable non-topological skyrmion solutions 
as well as topological ones in a background of charged black holes. 
The Skyrme model is valid at a greater scale than the quark 
confinement scale where the gauge fields can be assumed to be abelian. 
Although more complicated non-abelian monopole solutions are possible, 
depending on the details of the Higgs sector~\cite{brihaye01}, 
we shall restrict attention to the simplest case. 
The black hole mass for abelian monopoles has the lower bound 
\begin{eqnarray}
	M=\frac{p}{\sqrt{G}}=pM_{pl}\approx 2.54\times 10^{-7} \;\; {\rm Kg}\, , \label{}
\end{eqnarray}
where we denoted the magnetic charge $p\approx 11.7$ and the Planck mass 
$M_{pl}\approx 2.1768\times 10^{-8}$ Kg. We shall also examine the stability 
of our solutions. The main obstacle to proton decay 
around the monopole black hole is electric charge conservation. 
The black hole cannot swallow the proton whole because 
this would tip it over the extremal limit. 
It should come as no surprise, therefore, that the non-topological 
solutions are stable. The introduction of charged fermions, which 
can carry the electric charge away, will be also briefly discussed. 

This work was carried out with the collaboration of I. Moss and 
E. Winstanley.  

\subsection{Field Equations and Static Solutions}
The Lagrangian is based on a gauged version of the original 
Skyrme Lagrangian constructed by Callan and Witten~\cite{callan-witten84}. 
The natural extension to the charged $SU(2)$ chiral field is gauging 
the Skyrme model. For a gauge transformation 
\begin{eqnarray}
	U \rightarrow U + ie\alpha \, [Q,U]\; ,\;\;\;\;\;
	A_{\mu}\rightarrow A_{\mu}+\partial_{\mu}\alpha \, , \label{}
\end{eqnarray}
one can define the covariant derivative 
\begin{eqnarray}
	D_{\mu}U = \partial_{\mu}U-ieA_{\mu}\,[Q,U]  \label{}
\end{eqnarray}
where $A_{\mu}$ is the photon field, $e$ is the charge of proton 
electric charge in unrationalized units, and 
\begin{eqnarray}
	Q=\left( \begin{array}{lcr}
	 \frac{2}{3} & 0 \\
	 0 & -\frac{1}{3} \end{array} \right) \label{}
\end{eqnarray}
is the charge matrix of quarks. 
Replacing merely the derivatives in the Lagrangian with the 
covariant derivatives $D_{\mu}U=\partial_{\mu}U-ieA_{\mu}\,[Q,U]$ 
is, however, not sufficient in the sense that QCD anomalous processes, 
such as $\pi^{0}\rightarrow \gamma\gamma$ or the $\gamma \pi^{+}
\pi^{-}\pi^{0}$ vertex, are not included. 
This is a manifestation of the chiral symmetry breaking 
due to the presence of the $U(1)$ gauge fields. Thus the correct 
Lagrangian should include an anomalous term ${\cal L}_{A}$. 
The anomalous term arises from the Wess-Zumino term which vanishes 
in the $SU(2)$ Skyrme model. 
It has been worked out in Refs.~\cite{callan-witten84,witten83} that the 
gauge invariant form of the Wess-Zumino term $\Gamma (U)$ 
has additional terms given by  
\begin{eqnarray}
	&& {\hat \Gamma}(U,A_{\mu}) = \Gamma (U)+e\int d^{4}x\, A_{\mu}J^{\mu}
	-ie^{2}\int d^{4}x\,F_{\mu\nu}A_{\rho}T_{\sigma} \label{}
\end{eqnarray}
which reads 
\begin{eqnarray}
	{\cal L}_{A}=\frac{1}{16\pi^{2}}\epsilon^{\mu\nu\rho\sigma}
	[eA_{\mu}\,{\rm tr}\,(QL_{\nu}L_{\rho}L_{\sigma}-QR_{\nu}R_{\rho}R_{\sigma})
	-ie^{2}F_{\mu\nu}A_{\rho}T_{\sigma}] \label{}
\end{eqnarray}
where the left current $L_{\mu}=U^{\dagger}\partial_{\mu}U$ and right current 
$R_{\mu}=U\partial_{\mu}U^{\dagger}$ and 
\begin{eqnarray}
	T_{\sigma}={\rm tr}\,(Q^{2}L_{\sigma}-Q^{2}R_{\sigma}+\frac{1}{2}Q
	U^{\dagger}QUL_{\sigma}-\frac{1}{2}QUQU^{\dagger}R_{\sigma}) \label{}
\end{eqnarray}
have been defined. 
Consequently, we have the following gauged Einstein-Skyrme Lagrangian  
\begin{eqnarray}
	 {\cal L}={\cal L}_{S}+{\cal L}_{A}+{\cal L}_{em}+{\cal L}_{G} \label{mono-lag}
\end{eqnarray}
where 
\begin{eqnarray}
	{\cal L}_{S}&=&\frac{F_{\pi}^{2}}{16}{\rm tr}
	\,(U^{\dagger}D_{\mu}UU^{\dagger}D_{\mu}U)+\frac{1}{32a^{2}}{\rm tr}\,
	([U^{\dagger}D_{\mu}U,U^{\dagger}D_{\nu}U]^{2}) \nonumber \\
	{\cal L}_{A}&=&\frac{e}{16\pi^{2}}\epsilon^{\mu\nu\rho\sigma}
	A_{\mu}{\rm tr}\,[Q(U^{\dagger}\partial_{\nu}UU^{\dagger}
	\partial_{\rho}UU^{\dagger}\partial_{\sigma}U
	+\partial_{\nu}UU^{\dagger}\partial_{\rho}U
	U^{\dagger}\partial_{\sigma}UU^{\dagger})] \nonumber \\ 
	&& +\frac{ie^{2}}{8\pi^{2}}\epsilon^{\mu\nu\rho\sigma}
	(\partial_{\mu}A_{\nu})A_{\rho} \nonumber \\
	&& \times \, {\rm tr}\,[Q^{2}(\partial_{\sigma}U)U^{\dagger} 
	+Q^{2}U^{\dagger}(\partial_{\sigma}U)+\frac{1}{2}Q(\partial_{\sigma}U)
	QU^{\dagger}-\frac{1}{2}QUQ(\partial_{\sigma}U^{\dagger})] \nonumber 
\end{eqnarray}
and the free actions are 
\begin{eqnarray}
	{\cal L}_{em}=-\frac{1}{16\pi}F_{\mu\nu}F^{\mu\nu} , \;\;\;\;\;
	{\cal L}_{G}=\frac{1}{16\pi G}R \,. \nonumber \label{}
\end{eqnarray}
The gauge-invariant baryon current is then given by 
\begin{eqnarray}
	B^{\mu}&=&\frac{\epsilon^{\mu\nu\rho\sigma}}{24\pi^{2}}
	[{\rm tr}\,(L_{\nu}L_{\rho}L_{\sigma})+3ieA_{\nu}{\rm tr}\,
	Q(L_{\rho}L_{\sigma}-R_{\rho}R_{\sigma})
	+3ie\partial_{\nu}A_{\rho}{\rm tr}\,Q(L_{\sigma}-R_{\sigma})] 
	\nonumber \\
	&=& \frac{\epsilon^{\mu\nu\rho\sigma}}{24\pi^{2}}
	{\rm tr}\,(L_{\nu}L_{\rho}L_{\sigma})
	+\frac{\epsilon^{\mu\nu\rho\sigma}}{24\pi^{2}}
	\,\partial_{\nu}\,[3ieA_{\rho}{\rm tr}\,
	Q(L_{\sigma}-R_{\sigma})] \label{jb}
\end{eqnarray} 
which shows the additional current is a total divergence. 

In the spherically symmetric case with a magnetic charge the gauge field 
has the form
\begin{eqnarray}
	A=p\,(1-\cos\theta)\,d\phi +\Phi \, dt \label{gauge-field}
\end{eqnarray}
where $p$ is a magnetic charge and the Dirac quantization condition is 
$pe=1$. The usual skyrmion has a magnetic moment which would interact 
with a magnetic monopole and break spherical symmetry. We use instead 
a non-topological ansatz for the chiral field 
\begin{eqnarray}
	U=e^{if(r,t)\tau^{3}} \,.\label{non-topo-ansatz}
\end{eqnarray}
One can see that this field is made up of neutral pions 
and commutes with the charge matrix $Q=\frac{1}{6}
+\frac{1}{2}\tau^{3}$. 
Nevertheless it has a non-zero total electric and baryonic charge due to 
the effects of anomalies as we shall see below. 
After inserting the ansatz (\ref{non-topo-ansatz}) into 
the zeroth component of the baryonic current (\ref{jb}), one obtains
\begin{eqnarray}
	n_{B}=\frac{ep}{2\pi}[f(\infty)-f(0)]\,. \label{}
\end{eqnarray}
The solution with the boundary conditions $f(0)=0$ and $f(\infty)=2\pi$ 
possesses unit baryon number and hence it can be interpreted as a baryon 
surrounding the monopole. 

If the field ansatz (\ref{non-topo-ansatz}) is substituted into the 
meson and electromagnetic interaction terms in the Lagrangian, 
they become 
\begin{eqnarray}
	&& {\cal L}_{S}=-\frac{F_{\pi}^{2}}{8}(\partial f)^{2} \label{df} \\
	&& {\cal L}_{A}=-\frac{e^{2}}{8\pi^{2}}E_{i}B^{i}\,f \\
	&& {\cal L}_{em}=\frac{1}{8\pi}(E^{2}-B^{2}) \label{}
\end{eqnarray}
where the index $i=1,2,3$ and the electromagnetic fields $E_{i}$ and 
$B_{i}$ are defined by $F_{0i}=\sqrt{-g_{tt}}\,E_{i}$ and $F_{ij}=\epsilon_{ijk}B^{k}$.  
When combined 
\begin{eqnarray}
	{\cal L}_{A}+{\cal L}_{em}=\frac{1}{8\pi}\left(E-\frac{e^{2}}{2\pi}Bf\right)^{2}
	-\frac{1}{8\pi}B^{2}\left(1-\frac{e^{4}}{4\pi^{2}}f^{2}\right) \,.\label{factor}
\end{eqnarray}
The extrema of the action occur when the electric field is given by
\begin{eqnarray}
	E=\frac{e^{2}}{2\pi}Bf \,. \label{}
\end{eqnarray}
This situation is reminiscent of the factorization of the Lagrangian 
that occurs for a BPS monopole~\cite{bogomolnyi}.

The electric field implies a total charge 
\begin{eqnarray}
	q=\frac{e^{2}p}{2\pi}f \label{q-p}
\end{eqnarray}
or asymptotically 
\begin{eqnarray}
	q_{\infty}=n_{B}e  \label{}
\end{eqnarray}
and the $n_{B}=1$ solution can therefore be interpreted 
as a proton. If a black hole appears in the background, the inner 
boundary condition for the field $f$ should be imposed not at 
the origin but at the event horizon $r=r_{+}$. Thus the baryon number 
in the presence of an event horizon will be defined as 
\begin{eqnarray}
	n_{B}=\frac{ep}{2\pi}[f(\infty)-f(r_{+})]\, , \label{}
\end{eqnarray}
which implies the baryon number swallowed by the black hole is 
\begin{eqnarray}
	n^{-}_{B}=\frac{ep}{2\pi}[f(r_{+})-f(0)]\,. \label{}
\end{eqnarray}
Thus the total baryon number can be recovered as the sum 
\begin{eqnarray}
	n^{tot}_{B}=n_{B}+n_{B}^{-}\,. \label{}
\end{eqnarray}
If $f(r_{+})=0$, the baryon number is still an integer and conserved. 
This configuration represents a proton tightly bound to the black hole. 
On the other hand, if $f$ takes some positive value at the horizon 
the baryon number is not an integer and the skyrmion carries 
fractional baryon number and electric charge. 
This configuration will be interpreted as a proton partially 
swallowed by the black hole. In particular, $f(r_{+})=2\pi$ 
means that the black hole has swallowed a whole proton, leaving 
nothing outside the horizon. 

It is interesting to observe that, while the baryon number disappears 
inside the horizon, the electric charge of the black hole can still be 
measured outside, turning the monopole black hole into a dyon black hole. 
Therefore, while the baryon number conservation is violated, charge 
conservation is not violated. 

\subsubsection{Extremal black hole solutions}

In the extremal case we can obtain a general solution based on the 
Papapetrou-Majumdar metrics~\cite{majumdar,papapetrou}. 
We begin with the background metric fixed and later generalize to 
solve the full Einstein equations with chiral matter. 
The Papapetrou-Majumdar metrics have the form 
\begin{eqnarray}
	ds^{2}=-H^{2}dt^{2}+H^{2}(dx^{2}+dy^{2}+dz^{2}) \label{majumdar}
\end{eqnarray}
where 
\begin{eqnarray}
	H=1+\sum_{n=1}^{n_{M}}\frac{GM_{n}}{R_{n}} \label{}
\end{eqnarray}
and $R_{n}$ is the ordinary Euclidean distance from the point mass $M_{n}$ 
located in three-dimensional space. 
We also associate these point masses with magnetic charges $P_{n}=G^{1/2}M_{n}$, 
and the magnetic field 
\begin{eqnarray}
	B=G^{-1/2}H^{-1}\partial H \,. \label{}
\end{eqnarray}
The matter Lagrangian obtained earlier (\ref{df}) has the form 
\begin{eqnarray}
	{\cal L}=-\frac{1}{8}F_{\pi}^{2}\,(\partial f)^{2}
	-\frac{1}{8\pi} B^{2}\,(1+\alpha^{2}f^{2}) \label{df-b}
\end{eqnarray}
where we will set 
\begin{eqnarray}
	\alpha=\frac{e^{2}}{2\pi}\; ,\;\;\;\;\; \mu^{2}=\pi GF_{\pi}^{2}\,. \label{}
\end{eqnarray}
The Skyrme field equation obtained from the Lagrangian on this background 
becomes 
\begin{eqnarray}
	-\mu^{2}\partial^{2}f+\alpha^{2}H^{-2}(\partial H)^{2}f = 0\,. \label{}
\end{eqnarray}
We lose no generality by taking equal charges $P_{n}=p$. The solution with 
baryon number $n_{B}=n_{M}$ is then 
\begin{eqnarray}
	f=2\pi H^{-s} \label{}
\end{eqnarray}
where 
\begin{eqnarray}
	s=-\frac{1}{2}+\sqrt{\frac{1}{4}+\frac{\alpha^{2}}{\mu^{2}}}\,. \label{}
\end{eqnarray}
Since $F_{\pi}\ll m_{pl}$, we can use $s \approx \alpha/\mu$ for the pion model. 

For a single black hole, the Reissner-Nordstr{\"o}m coordinate $r$ is 
related to $R$ by $R=r-r_{+}$ and we obtain 
\begin{eqnarray}
	f=2\pi \left(1-\frac{r_{+}}{r}\right)^{s}\,. \label{f-s}
\end{eqnarray}
The field is effectively expelled by the black hole and vanishes on the horizon 
$r=r_{+}$. 

The mass of the chiral field configuration can be obtained by 
integrating the Lagrangian (\ref{df-b}), 
\begin{eqnarray}
	m_{f}=\frac{1}{8}F_{\pi}^{2}\int_{\Sigma} f\partial_{i}f\, dS^{i} \label{}
\end{eqnarray}
where $\Sigma$ is a large surface containing all of the masses. This gives 
\begin{eqnarray}
	m_{f}=2\pi^{3}sF_{\pi}^{2}G\sum_{n}M_{n} \approx \pi^{3/2}n_{B}
	\,e\,F_{\pi}\,. \label{}
\end{eqnarray}
The total mass in the chiral field is much less than one baryon mass per mass point. 
We can see how it is energetically favorable for a free skyrmion to 
change its internal configuration from the original Skyrme form to the simpler 
form used here when it comes into contact with a black hole monopole. 
The total topological description of this transformation for a single 
monopole is exactly as described in \cite{callan-witten84}. 

It is interesting to see that the electrostatic energy cancels due to 
the factorization occurring in the Lagrangian (\ref{factor}). The chiral field 
mass is independent of the separation of the holes and therefore there are 
no forces between them. This is similar to the situation for BPS monopole 
solutions~\cite{bogomolnyi}, and suggests that there is a solution of the 
full Einstein-matter system. This existence of the solution will now be 
demonstrated. 

The spatial part of the Einstein tensor for the metric (\ref{majumdar}) is 
\begin{eqnarray}
	G_{ij}=-2H^{-2}(\partial_{i}H)(\partial_{j}H)+H^{-2}(\partial H)^{2}
	g_{ij} \label{}
\end{eqnarray}
and the Ricci scalar is 
\begin{eqnarray}
	R=-2H^{-1}\partial^{2}H \,. \label{}
\end{eqnarray}
Substituting the Einstein tensor for the Lagrangian (\ref{df-b}) into 
the Einstein equations gives 
\begin{eqnarray}
	&& H^{-1}\partial^{2}H=-\mu^{2}(\partial f)^{2} \\
	&& H^{-2}(\partial_{i}H)(\partial_{j}H)=-\mu^{2}(\partial_{i}f)
	(\partial_{j}f)+GB_{i}B_{j}(1+\alpha^{2}f^{2})\,. \label{}
\end{eqnarray}
These make up a complete system of equations when we include the 
Maxwell equation
\begin{eqnarray}
	\partial (HB)=0\,. \label{}
\end{eqnarray}
The second Einstein equation implies that $\partial f$, $\partial H$ 
and $B$ are all parallel. We therefore impose a condition 
$f\equiv f(u)$, $B=b(u)H^{-1}\partial H$, where 
\begin{eqnarray}
	u=-\mu^{-1}{\rm log}H\,. \label{u-mu}
\end{eqnarray}
The system of equations becomes equivalent to an ordinary 
differential equation with independent parameter $u$, 
\begin{eqnarray}
	f''+\mu (1+f'^{2})f'-\alpha^{2}b^{2}f=0 \label{f-alpha}
\end{eqnarray}
where 
\begin{eqnarray}
	b^{2}=\frac{1+f'^{2}}{1+\alpha^{2}f^{2}}\,. \label{}
\end{eqnarray}
The horizon corresponds to $u\rightarrow -\infty$ and the 
far region to $u\rightarrow 0$. The horizon must therefore 
be at a critical point of the first-order system corresponding 
to (\ref{f-alpha}). There is only one critical point, $(f,f')=(0,0)$, 
hence 
\begin{eqnarray}
	(f,f')\rightarrow (0,0) \;\;\;\;\; {\rm as} \;\;\; 
	u\rightarrow -\infty\,. \label{}
\end{eqnarray}
Since the critical point is a saddle, the solution is unique 
and exists for all values of $\mu$. Having obtained the unique 
solution to (\ref{f-alpha}), we then define 
\begin{eqnarray}
	V(u)=1+\mu \int_{u}^{0}b(x)e^{-\mu x}dx\,. \label{v-mu}
\end{eqnarray}
It is easily seen from (\ref{u-mu}) that 
\begin{eqnarray}
	\partial_{i}V=V'\partial_{i}u =HB\,.  \label{}
\end{eqnarray}
Hence the Maxwell equation implies $\partial^{2}V=0$ 
and we can write 
\begin{eqnarray}
	V=1+\sum_{n}\frac{GM_{n}}{R_{n}}\,. \label{}
\end{eqnarray}
Inverting (\ref{v-mu}) gives $u(V)$. 

\subsubsection{Spherically symmetric solutions}

In the non-extremal case we shall consider spherically 
symmetric metrics which can be parameterized in the form
\begin{eqnarray}
	ds^{2}=-\frac{\Delta}{r^{2}}e^{2\delta}dt^{2}+\frac{r^{2}}{\Delta}dr^{2}
	+r^{2}d\Omega^{2} \label{}
\end{eqnarray}
where $\Delta$ and $\delta$ are functions of $r$ and $t$. After inserting 
the metric and the other field ansatz (\ref{gauge-field}) and (\ref{non-topo-ansatz}) 
into the Einstein field equations, one obtains 
\begin{eqnarray}
	&& (\Delta e^{\delta}f')'-\frac{\lambda^{2}}{r^{2}}e^{\delta}f
	=-\frac{2r^{4}}{\Delta^{3}}e^{-\delta}{\dot \Delta}{\dot f}
	-\frac{r^{4}}{\Delta}e^{-\delta}{\dot \delta}{\dot f}
	+\frac{r^{4}}{\Delta^{2}}e^{-\delta}{\ddot f} \label{sphe-sky} \\ 
	&& \delta'=\mu^{2}r\left(\frac{r^{4}}{\Delta}e^{-2\delta}{\dot f}^{2} 
	+f'^{2}\right) \label{sphe-del}\\
	&& e^{-\delta}\left(\frac{\Delta e^{\delta}}{r}\right)'= 
	1-\frac{\mu^{2}\lambda^{2}}{r^{2}}f^{2}-\frac{Gp^{2}}{r^{2}} \label{sphe-e}
\end{eqnarray}
where $\mu$ and $\lambda$ are constants, 
\begin{eqnarray}
	\mu^{2}=\pi F_{\pi}^{2}G \;,\;\;\; \lambda^{2}=
	\frac{e^{4}p^{2}}{4\pi^{3}F_{\pi}^{2}}\,. \label{}
\end{eqnarray}
The electric charge within a sphere of radius $r$ is given by equation (\ref{q-p}). 

For very small $\mu$, which is the case for pions, the chiral field has little 
effect on the background metric and we may take $\delta=0$ and express $\Delta$ 
in terms of the mass $M$, electric charge $Q$ and magnetic charge $p$ of the 
black hole as 
\begin{eqnarray}
	\Delta =r^{2}-2GMr+G(Q^{2}+p^{2})\,. \label{}
\end{eqnarray}
The Skyrme field equation (\ref{sphe-sky}) on this background is therefore 
\begin{eqnarray}
	(\Delta f')'-\frac{\lambda^{2}}{r^{2}}f=0\,.  \label{f-lambda}
\end{eqnarray}
This should be solved subject to the boundary condition on $f_{\infty}$ 
which fixes the total charge, 
\begin{eqnarray}
	q_{\infty}=\frac{e^{2}p}{2\pi}f_{\infty}=n_{B}e\,. \label{}
\end{eqnarray}
The non-extremal black hole possesses two horizons at $r=r_{-}$ and 
$r=r_{+}$ $(r_{+}>r_{-})$, related to the mass and charge by 
\begin{eqnarray}
	GM=\frac{1}{2}(r_{-}+r_{+})\;,\;\;\; GQ^{2}=r_{-}r_{+}-Gp^{2}\,. \label{m-q}
\end{eqnarray}
The solution to Eq.~(\ref{f-lambda}) can be obtained analytically, 
\begin{eqnarray}
	f=2\pi n_{B}\frac{P_{q}((r_{+}+r_{-})/(r_{+}-r_{-})
	-[2r_{+}r_{-}/(r_{+}-r_{-})]/r)}
	{P_{s}((r_{+}+r_{-})/(r_{+}-r_{-}))} \label{}
\end{eqnarray}
where $P_{s}(z)$ is a Legendre function and 
\begin{eqnarray}
	s=-\frac{1}{2}+\sqrt{\frac{1}{4}+\frac{\lambda^{2}}{r_{+}r_{-}}}\,. \label{}
\end{eqnarray}
The black hole become a dyon with electric charge related to the value 
of $f$ at the event horizon, 
\begin{eqnarray}
	Q=\frac{n_{B}e}{P_{s}((r_{+}+r_{-})/(r_{+}-r_{-}))}\,.  \label{}
\end{eqnarray}
This relation can be solved, together with (\ref{m-q}), to obtain $Q\equiv Q(M)$, 
showing the existence of a one-parameter family of solutions ($n_{B}$ and $p$ 
being regarded as fixed). In particular, $Q\rightarrow 0$ as $M$ approaches the 
extremal limit $pM_{pl}$ and the meson field is expelled from the hole. 

Larger values of $\mu$ may be realized for a hypothetical model where 
$U$ is unrelated to pions, and this is discussed below. 
We will consider a static solution. We can replace $\Delta$ by a mass 
function $m(r)$, defined implicitly by the relation 
\begin{eqnarray}
	\Delta =r^{2}-2Gmr+G(p^{2}+q^{2}) \label{}
\end{eqnarray}
where the charge is given by equation (\ref{q-p}). The static 
equations become 
\begin{eqnarray}
	&& m'=\frac{\Delta}{2Gr}\delta'+\mu^{2}\lambda^{2}ff' \\
	&& \delta'=\mu^{2}r(f')^{2} \\
	&& f''+\left(\frac{\Delta'}{\Delta}+\delta'\right)f'
	-\frac{\lambda^{2}}{r^{2}\Delta}f=0\,. \label{ddf-delta}
\end{eqnarray}
Suitable boundary conditions are $f\rightarrow 2\pi$ and $\delta\rightarrow 0$ 
as $r\rightarrow \infty$. 

In the numerical results the fields are scaled to the horizon size, 
\begin{eqnarray}
	{\hat r}=\frac{r}{r_{+}}\;,\;\;\; {\hat m}=\frac{Gm}{r_{+}}\,. \label{}
\end{eqnarray}
The solutions are parameterized by a parameter ${\hat p}$, defined by
\begin{eqnarray}
	{\hat p}^{2}=\frac{Gp^{2}}{r_{+}^{2}} \label{}
\end{eqnarray}
which is restricted to ${\hat p}\leqslant 1$.

\begin{figure}
\hspace{2cm}
\includegraphics[height=8cm, width=11cm]{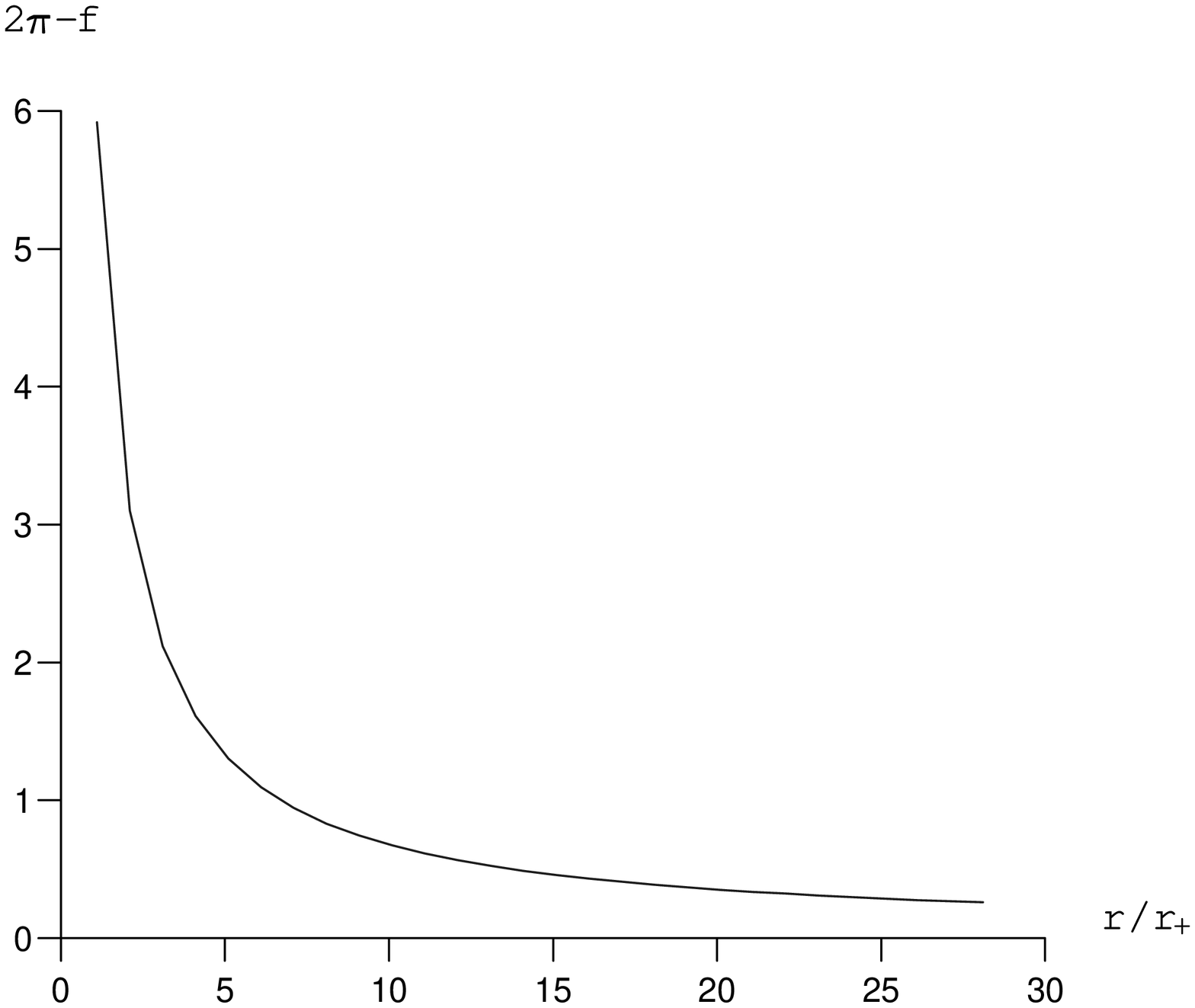}
\caption{\label{fig1} Profile function $f$ as a function of 
${\hat r}=r/r_{+}$ for an extremal hole and $\mu =10^{-4}$.}
\hspace{2cm}
\includegraphics[height=8cm, width=11cm]{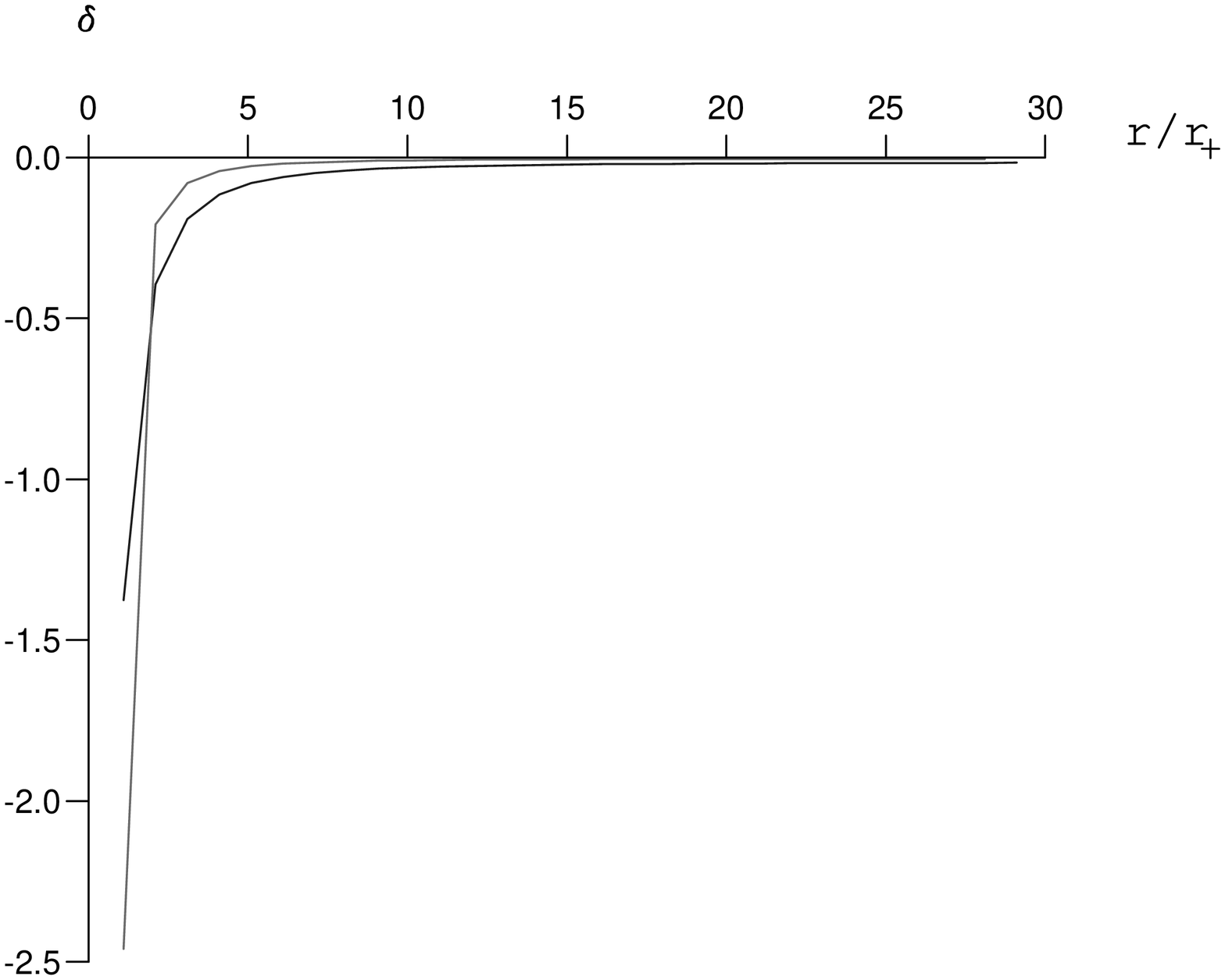}
\caption{\label{fig2} Backreaction $\delta$ $(\times 10^{3})$ as a function of $\hat{r}=r/r_+$ for a non-extremal black hole, 
${\hat p} =pm_{pl}/r_+=  0.9$. Results for $\mu=10^{-3}$ and $\mu=10^{-4}$ are shown.}
\end{figure}

The extremal black hole solutions have $\Delta_{+}=\Delta_{+}'=0$. 
The regular solution to equation (\ref{ddf-delta}) has 
\begin{eqnarray}
	f_{+}=0\;,\;\;\; Q=\frac{e}{2\pi}f_{+}=0\;,\;\;\; {\hat p}=1\,. \label{}
\end{eqnarray}
Hence the proton lies fully outside the black hole, as we saw before. 
The numerical solution for $f$ is shown in Fig.~\ref{fig1}. This agrees with 
well with the result (\ref{f-s}), because the value of $\mu$ used here is still 
quite small. The results are still qualitatively similar for chiral models with 
$\mu$ of order one. 

For the non-extremal solution, we begin the integration of the field equations 
close to the horizon, with 
\begin{eqnarray}
	&& {\hat m}={\hat m}_{0}+{\hat m}_{1}({\hat r}-1)+{\hat m}_{2}
	({\hat r}-1)^{2}+\cdots \\
	&& \delta =\delta_{+}+\delta_{1}({\hat r}-1)+\cdots \\
	&& f=f_{+}+f_{1}({\hat r}-1)+\cdots \label{}
\end{eqnarray}
where $\delta_{+}$ and $f_{+}$ are shooting parameters determined so as to 
satisfy the boundary conditions at infinity. 

\begin{figure}
\hspace{2cm}
\includegraphics[height=8cm, width=11cm]{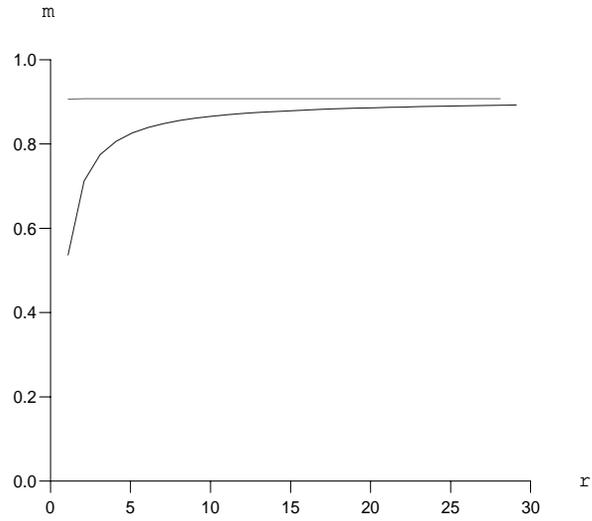}
\caption{\label{fig3} Mass function $m$ as a function of $\hat{r}$ 
         for  ${\hat p} = 0.9$. Results for $\mu=10^{-3}$ and $\mu=10^{-4}$ are shown}
\end{figure}

As can be seen from the above expansion, the Skyrme 
field must have a nonzero value at the horizon 
otherwise the only allowed solution is the trivial one. 
Consequently the nonextremal black hole acquires 
an electric charge 
\begin{equation}
	Q = \frac{{\rm e}}{2\pi}f_{+},
\end{equation}
and allows the skyrmion to have fractional electric charge.  
The numerical results for this solution are shown in 
Figs.~\ref{fig2} and \ref{fig3}.   
Again, these agree well with the fixed 
background for small values of $\mu$. 

\begin{figure}
\hspace{2cm}
\includegraphics[height=8cm, width=11cm]{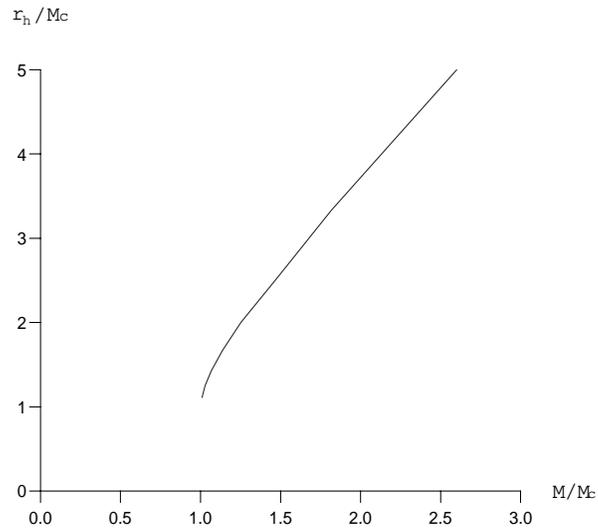}
\caption{\label{fig4} Horizon radius $r_{+}/M_{c}$ as a function 
          of the black hole  mass $M/M_{c}$ for $\mu =10^{-4}$.}
\end{figure}

\begin{figure}
\hspace{2cm}
\includegraphics[height=8cm, width=11cm]{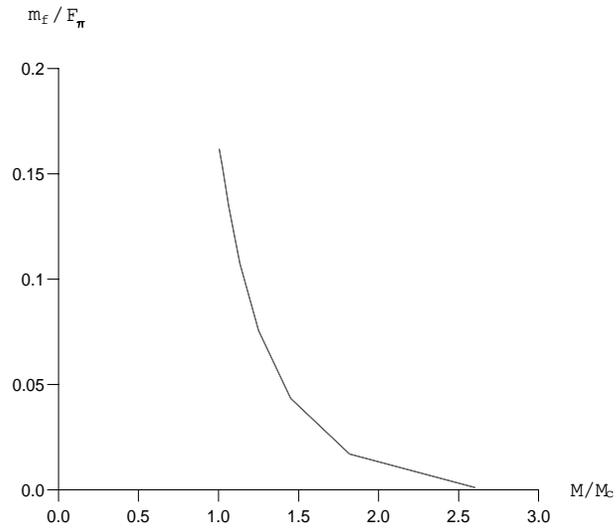}
\caption{\label{fig5} Skyrmion mass $m_f$ as a function 
          of the black hole mass $M/M_{c}$ for $\mu =10^{-4}$.}
\end{figure}

We have a single one parameter family of solutions
with ${\hat p}\le 1$. In Figs.~\ref{fig4} and \ref{fig5}, we plot 
the horizon radius $r_{+}$ and 
skyrmion mass $m_{f}$ as functions of black hole mass $M$.
Figure \ref{fig4} is related to the entropy of the
black hole. The entropy of the black hole can be defined 
in the form 
\begin{eqnarray}
	S=\frac{1}{4G} {\cal A}_{bh} \label{}
\end{eqnarray}
where ${\cal A}_{bh}$ is the area of black holes, which is ($\pi r_+^2$). 
The research in the entropy of black holes has revealed 
that the entropy of an extremal Reissner-Nordstrom black hole 
is zero, despite its finite size and it cannot be reached from the 
non-extremal state~\cite{hawking95}. Thus the extremal black hole 
is a thermodynamically different object from the non-extremal one. 

\begin{figure}
\hspace{2cm}
\includegraphics[height=8cm, width=11cm]{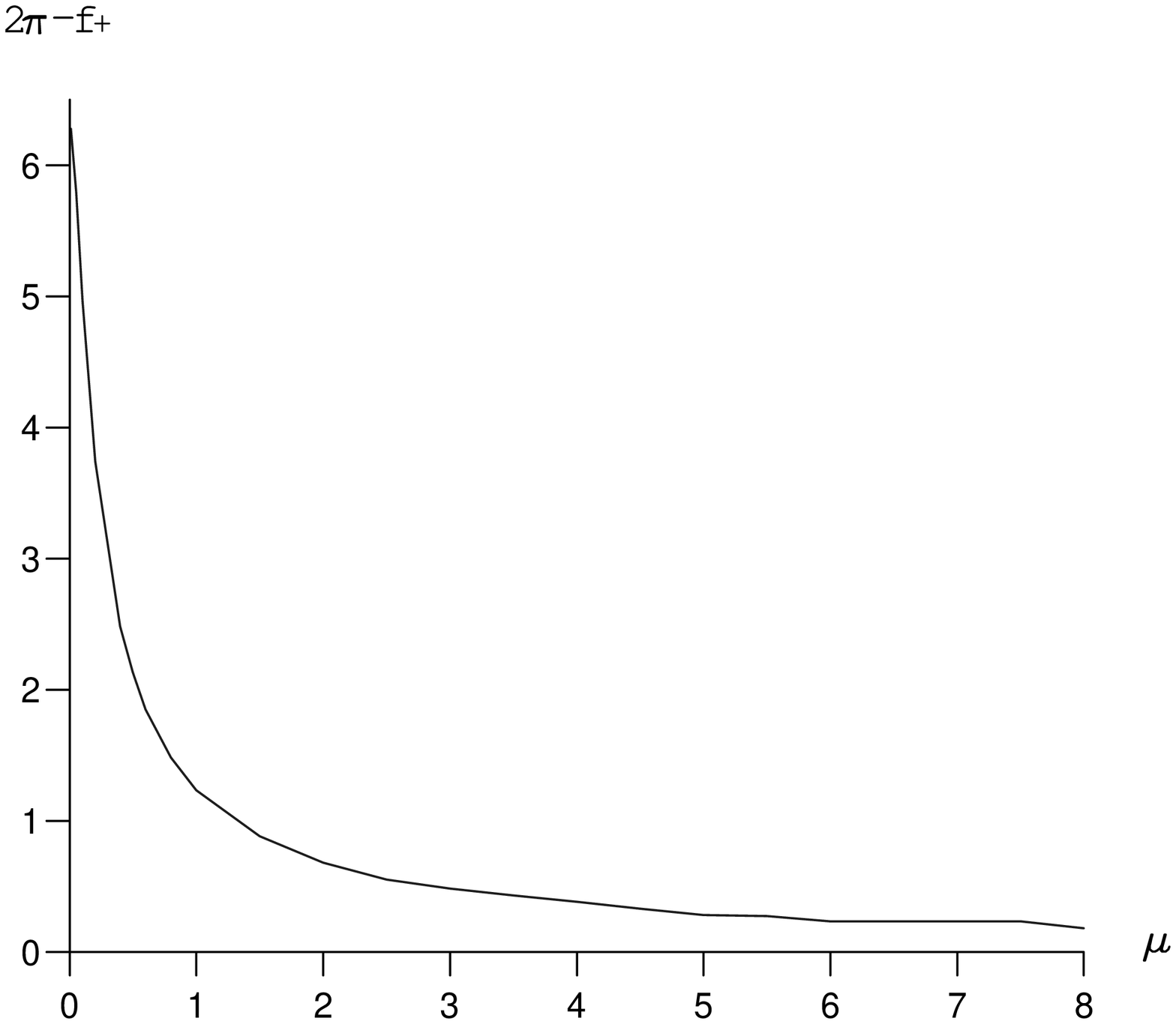}
\caption{The value of $f$ at the horizon for various values of $\mu$.}
	\protect\label{fig6}
\end{figure}

The other figure shows how the 
proportion of the skyrmion 
which is swallowed by the black hole increases with the 
black hole mass.

Figure \ref{fig6} shows how the horizon value of 
$f$ changes as the coupling constant $\mu$ changes. 
As can be seen from the figure, for small $\mu$ 
the electromagnetic interaction 
is dominant so that the skyrmion is absorbed 
by the black hole to a lesser extent. On the other hand 
for large $\mu$ the gravitational interaction is 
dominant so that most of the skyrmion is absorbed by 
the black hole. 

\pagebreak

 \subsection{Stability Analysis}

In this section we show that the skyrmion solutions 
which we have obtained are stable under spherically
symmetric linear perturbations. We shall begin with 
the analysis of a skyrmion on the fixed background.

In the fixed-background case the skyrme field is the only perturbed 
field and can be expanded about the skyrmion 
solution $f_{0}$ by writing
\begin{equation}
	f(r,t) = f_{0}(r) + e^{i\omega t}\xi(r) . 
	\label{pertf}
\end{equation}
Equation (\ref{pertf}) is inserted into equation (\ref{sphe-sky}) 
with $\delta=0$ to obtain the eigenvalue equation
\begin{equation}
	-\left(\Delta_{0}\xi^{\prime }\right)^{\prime }
         +\frac{\lambda^{2}}{r^{2}}\xi
	= \frac{r^{4}}{\Delta_{0}}\omega^{2}\xi ,
	\label{fixeigen}
\end{equation}
where the background equation has been used.

If $\omega $ is real and $\omega^{2} > 0$ the solution is stable, 
and if $\omega $ is imaginary and $\omega^{2} < 0$ 
it is unstable since the mode can grow 
or decay exponentially under the small perturbation.
To show which is the case we multiply both sides of 
equation (\ref{fixeigen}) by $\xi$ 
and integrate in $r$ from the horizon to infinity 
\begin{equation}
	\int_{r_{+}}^{\infty}\; \left[\frac{\Delta_{0}}{2} 
        \xi^{\prime 2}
	+\frac{\lambda^{2}}{r^{2}}\xi^{2}\right]\;dr = 
         \omega^{2}\int_{r_{+}}^{\infty} \; 
	\frac{r^{4}}{\Delta_{0}}\xi^{2} \; dr \; ,
	{}
\end{equation}
where integration by parts and boundary conditions have 
been used.
It can be seen that the integrands of both sides are positive definite, 
which means 
that $\omega^{2} >0$. Hence the skyrmion on the fixed background 
is linearly stable .

Next we analyze the stability of the skyrmion with backreaction.
In this case we have to expand the metric as well as the skyrme 
field around the classical solutions $f_{0}$, $ \delta_{0}$ and 
$ \Delta_{0}$
\begin{eqnarray*}
	f(r,t) & = & f_{0}(r)+f_{1}(r,t)  \\
	\delta(r,t) & = & \delta_{0}+\delta_{1}(r,t)  \\
	\Delta(r,t) & = & \Delta_{0}+\Delta_{1}(r,t)  \; .
\end{eqnarray*}
These are substituted into Eqs.~(\ref{sphe-sky})-(\ref{sphe-e}) 
to obtain the following 
coupled equations up to first order
\begin{eqnarray}
	&& \left[\left(\Delta_{0}\delta_{1}f_{0}^{\prime }
         +\Delta_{0}f_{1}^{\prime }
	+\Delta_{1}f_{0}^{\prime }\right)e^{\delta_{0}}
         \right]^{\prime }-\frac{\lambda^{2}}
         {r^{2}}\left(\delta_{1}f_{0}
	+f_{1}\right)e^{\delta_{0}}=
	\frac{r^{4}}{\Delta_{0}}e^{-\delta_{0}}\ddot{f}_{1}
	\label{pf} \\
	&& \delta_{1}^{\prime }=2\mu^{2}
          rf_{0}^{\prime }f_{1}^{\prime }
	\label{del} \\
	&& \left(\frac{2\mu^2\lambda^{2}}{r^{2}}f_{0}f_{1}
	 +\frac{\Delta_{0}}{r}\delta_{1}^{\prime }
           \right)e^{\delta_{0}}= -\left(\frac{\Delta_{1}}{r}
          e^{\delta_{0}}\right)^{\prime } .
	\label{mass}
\end{eqnarray}
Equation (\ref{mass}) can be integrated with the help 
of the static field equation,
\begin{equation}
	\Delta_{1}=-2\mu^{2}r\Delta_{0}f_{0}^{\prime }
         f_{1} \; .
	\label{intmass}
\end{equation}
Substituting Eqs.~(\ref{del}) and (\ref{intmass}) 
into Eq.~(\ref{pf}) one obtains the first order equation
for $f_{1}$ 
\begin{equation}
	\left(\Delta_{0}e^{\delta_{0}}f_{1}^{\prime }
          \right)^{\prime }
	-\left[2\mu^{2}\left(r\Delta_{0}e^{\delta_{0}}
         f_{0}^{\prime 2}\right)^{\prime }
        +\frac{\lambda^{2}(M_{c})^{2}}{r^{2}}
	e^{\delta_{0}}\right]f_{1} = 
	\frac{r^{4}}{\Delta_{0}}e^{-\delta_{0}}\ddot{f}_{1} \; .
	{}
\end{equation}
Setting $f_{1}(r,t)=\xi (r)e^{i\omega t}$ one obtains an 
eigenvalue equation for $\xi$,
\begin{equation}
	-\left(\Delta_{0}e^{\delta_{0}}
        \xi^{\prime }\right)^{\prime }
	+\left[2\mu^{2}\left(r\Delta_{0}
         e^{\delta_{0}}f_{0}^{\prime 
	2}\right)^{\prime }+\frac{\lambda^{2}
         (M_{c})^{2}}{r^{2}}
	e^{\delta_{0}}\right]\xi = \omega^{2}
	\frac{r^{4}}{\Delta_{0}}e^{-\delta_{0}}\xi \; .
	\label{backeigen}
\end{equation}
We introduce the tortoise coordinate $r^{*}$ such that
\begin{equation}
	\frac{dr^{*}}{dr} = \frac{1}{\Delta_{0}e^{\delta_{0}}}  
	{}
\end{equation}
and $r^{* }$ runs from $-\infty$ to $+\infty$ as $r$ 
runs from $r_{+}$ to $+\infty$.
Then Eq.~(\ref{backeigen}) is reduced to 
the Sturm-Liouville equation
\begin{equation}
	-\frac{d^{2}\xi}{dr^{* 2}} + U\xi = 
        \omega^{2}r^{4}\xi  ,
	\label{eigen}
\end{equation}
where 
\begin{equation}
	U=\left[2\mu^{2}\left(r\Delta_{0}e^{\delta_{0}}
       f_{0}^{\prime 2}\right)^{\prime }+\frac{\lambda^{2}}{r^{2}}
	e^{\delta_{0}}\right]\Delta_{0}e^{\delta_{0}}  .
	{}
\end{equation}
On the left-hand side we have 
$r^{4}$, which makes the equation different from 
the previous eigenvalue equation with the fixed background.
However, since $r^{4}$ remains positive through the whole space, 
the same conditions for stability as the ordinary eigenvalue 
equation can be applied.
As can be seen by examining $U$, 
$U\to 0$ as $r\to r_+$, (i.e. $r^*\to-\infty$), $U\to U_\infty>0$ 
as $r\to\infty$, and $U>0$ in between. 
In addition the solution does not change its shape for any 
value of the coupling constant $\mu$. 
Therefore we can safely conclude that a skyrmion 
with backreaction is also linearly stable.

When the solutions are stable, proton decay can only take place
when extra particle fields are included in the model to carry the 
electric charge away.  
Since the underlying $SU(5)$ theory does not require  
lepton charge conservation, it should satisfy $\Delta B=\Delta L$. 
The configuration we concern here is spherically symmetric and thus 
the theory is reduced to one space and one time dimension. 
In this case the bosonization technique can be applied in order 
to include leptons~\cite{coleman75,mandelstam75}. 
According the technique, we simply replace fermionic current 
to a real scalar field as 
\begin{eqnarray}
	j^{\mu}={\bar \chi}(r,t)\gamma^{\mu}\chi (r,t)
	\rightarrow -\frac{1}{\sqrt{\pi}}
	\epsilon^{\mu\nu}\partial_{\nu}\phi \,(r,t) \,.\label{}
\end{eqnarray}
Then following the procedure described in Ref.~\cite{callan-witten84}, 
we obtain the equations for the fixed background (approximately)
\begin{eqnarray}
	\left(\Delta f^{\prime }\right)^{\prime } - 
	{\lambda^2\over r^{2}}(f-\phi)=-{r^4\over \Delta}\ddot f\, ,\\
\left(\Delta \phi^{\prime }\right)^{\prime } +
	{\lambda^2\over r^{2}}(f-\phi)=-{r^4\over \Delta}\ddot\phi \, ,
\end{eqnarray}
The stability arguments no longer apply. The dynamical process of a black hole 
swallowing a proton can be examined by solving these time-dependent 
field equations numerically. 
For a flat-space time, these were solved by Chemtob in Ref.~\cite{chemtob89} 
where the catalysis cross section was estimated approximately 
to be $1 {\rm mb}/\beta$ with velocity $\beta$, which confirms that 
the monopole catalysis proceeds at a strong-interaction scale without any 
suppression.    

\pagebreak 

\section{$B=2$ Black Hole Skyrmions}

In this section black hole solutions of $B=2$ skyrmions 
with axisymmetry are studied. 
The study by Hartmann {\it et al.} showed the Einstein-Yang-Mills-Higgs 
theory possesses axially symmetric monopole and black hole 
solutions~\cite{kleihaus97,hartmann02}. 
We follow their numerical technique to solve the Einstein-Skyrme model 
with axisymmetry. 
The obtained solution exhibits a torus shape in the energy 
and baryon density with a black hole at the center. 
Similarly to the case of $B=1$, the baryon number is not 
conserved and takes fractional values outside the horizon. 

Recent studies of theories with large extra dimensions indicate 
that a true Planck scale is of order a TeV and the production rate of black holes 
massive than the Planck scale become quite large~\cite{hamed98,antoniadis98}. 
Therefore, if a skyrmion represents a proton, this kind of black hole may 
be created in p-p collisions at the LHC in future. 

This work was carried out with the collaboration of T. Torii and K. Maeda. 

\subsection{Axially Symmetric Configurations}
 
The Lagrangian for the Einstein-Skyrme system has been already 
given in Eq.~(\ref{action}). 
Since the $B=2$ skyrmion is axially symmetric, we impose  
axially symmetric ansatz for the metric and chiral fields. 
We use the ansatz for the metric given in Ref.~\cite{kleihaus97}
\begin{eqnarray}
	ds^2 = -fdt^2 + \frac{m}{f}(dr^2 + r^2 d\theta^2) + \frac{l}{f} 
	r^2 \sin^2\theta d\varphi^2
\end{eqnarray}
where $f=f(r,\theta), \,\, m=m(r,\theta), \,\,{\rm and}\,\, l=l(r,\theta)$. 

The axially symmetric chiral fields can be parameterized by  
\begin{eqnarray}
	U=\cos F(r,\theta)+i{\vec \tau}\cdot{\vec n}_R\sin F(r,\theta)
	\label{chiral}
\end{eqnarray}
with ${\vec n}_R = (\sin\Theta\cos n\varphi, \sin\Theta\sin n\varphi, \cos\Theta)$ 
and $\Theta =\Theta (r,\theta )$.
The integer $n$ corresponds to the winding number of the Skyrme fields and for $B=2$ we have $n=2$.

In terms of $F$ and $\Theta$, the Lagrangian takes the form 
\begin{eqnarray}
	{\cal L}_S={\cal L}_{S}^{(1)}+{\cal L}_{S}^{(2)} \label{}
\end{eqnarray}
where 
\begin{eqnarray*}
	{\cal L}_S^{(1)}& =& -\frac{F_\pi^2}{8}\frac{f}{m}\left[(\partial_{r}F)^2
	+\frac{1}{r^{2}}(\partial_{\theta}F)^2+\left\{(\partial_{r}\Theta)^2+\frac{1}{r^{2}}
	(\partial_{\theta}\Theta)^2\right\}\sin^{2}F \right. \\
	&& \left. + \frac{n^{2}}{r^2\sin^2 \theta}\frac{m}{l}\sin^{2}{\Theta}\sin^{2}F \right] \\
	{\cal L}_S^{(2)}&=&-\frac{1}{2a^{2}r^{2}}\left(\frac{f}{m}\right)^{2}
	\biggl[\left(\partial_{[r}F\partial_{\theta]}\Theta\right)^{2} 
	+\frac{n^{2}}{\sin^{2}\theta}\frac{m}{l}\left\{(\partial_{r}F)^{2}
	+\frac{1}{r^{2}}(\partial_{\theta}F)^{2}\right\} \sin^{2}\Theta \\
	&&+ \frac{n^{2}}{\sin^{2}\theta}\frac{m}{l}
	\left\{(\partial_{r}\Theta)^{2}+\frac{1}{r^{2}}(\partial_{\theta}\Theta)^{2}
	\right\}\sin^{2}F\sin^{2}\Theta \biggr]\sin^{2}F  \label{}
\end{eqnarray*}
and we have defined the notation 
\begin{eqnarray}
	\partial_{[r}F\partial_{\theta]}\Theta =\partial_{r}F\partial_{\theta}\Theta
	-\partial_{\theta}F\partial_{r}\Theta \,. \label{}
\end{eqnarray}
The baryon current in curved spacetime is defined in Eq.~(\ref{baryon_current}).  
The baryon number is then given by integrating $B^{0}$ 
over the hypersurface $t=0$, 
\begin{eqnarray}
	B&=& \int drd\theta d\varphi \sqrt{g^{(3)}}\;B^{0}  \nonumber \\
	&=& -\frac{1}{\pi}\int drd\theta\;(\partial_{[r}F
	\partial_{\theta]}\Theta)
	\sin\Theta(1-\cos 2F) \nonumber \\ 
	&=& -\frac{1}{\pi}\int dFd\Theta \sin\Theta(1-\cos 2F) \nonumber \\ 
	&=& \left. \frac{1}{2\pi}(2F-\sin 2F)\cos \Theta \right|_{F_{0},
	\Theta_{0}}^{F_{1},\Theta_{1}} \label{}
\end{eqnarray}
where $(F_{0},\Theta_{0})$ and $(F_{1},\Theta_{1})$ are the values at the inner 
and outer boundary respectively. 
In flat spacetime we have
\begin{eqnarray*}
	(F_{0},\Theta_{0})=(\pi,0)\;\; {\rm and}\;\; (F_{1},\Theta_{1})=(0,\pi) , \label{}
\end{eqnarray*}
which gives $B=2$.  
In the presence of a black hole, the integration should be 
performed from the horizon to infinity, which change the values 
of $F_{0}$ and allow the $B$ to take fractional values of less 
than two. This situation can be interpreted as the black hole 
absorbing a skyrmion as was seen in the $B=1$ case. 

The energy density is given by the zero-zero component of the 
stress-energy tensor  
\begin{eqnarray}
	-T_{0}^{0}&=& \frac{F_{\pi}^{2}}{8}\frac{f}{m}\left[(\partial_{r}F)^{2}
	+\frac{1}{r^{2}}(\partial_{\theta}F)^{2}+\left\{(\partial_{r}
	\Theta)^{2}+\frac{1}{r^{2}}(\partial_{\theta}\Theta)^{2}\right\}
	\sin^{2}F \right. \nonumber \\
	&& \left. +\frac{n^{2}}{r^{2}\sin^{2}\theta}\frac{m}{l}\sin^{2}F
	\sin^{2}\Theta\right] +\frac{1}{2a^{2}r^{2}}\frac{f^{2}}{m^{2}}
	\left[\frac{}{}(\partial_{[r}F\partial_{\theta]}\Theta)^{2} \right. \nonumber \\
	&& \left. +\frac{n^{2}}{\sin^{2}\theta}\frac{m}{l}
	\left\{(\partial_{r}F)^{2}+\frac{1}{r^{2}}(\partial_{\theta}F)^{2}
	\right\}\sin^{2}\Theta \right. \nonumber \\
	&& \left. +\frac{n^{2}}{\sin^{2}\theta}\frac{m}{l}
	\left\{(\partial_{r}\Theta)^{2}+\frac{1}{r^{2}}(\partial_{\theta}\Theta)^{2}
	\right\}\sin^{2}F\sin^{2}\Theta\right]\sin^{2}F \, .  
\end{eqnarray}

\subsection{Boundary Conditions}
Let us consider the boundary conditions for the chiral fields and metric functions. 
At the horizon $r=r_{h}$, the zero-zero component of the metric satisfies 
\begin{eqnarray}
	g_{00}=-f(r_{h},\theta)=0 \, . \label{}
\end{eqnarray}
Regularity of the metric at the horizon requires  
\begin{eqnarray}
	m\,(r_{h},\theta)=l\,(r_{h},\theta)=0 \, . \label{}
\end{eqnarray}
The boundary conditions for $F(r,\theta)$ and $\Theta(r,\theta)$ at the horizon 
are obtained by expanding them at the horizon and inserting into the field equations 
which are derived from $\delta {\cal L}_{S}/\delta F=0$ and $\delta {\cal L}_{S}/\delta 
\Theta=0$ respectively. Consequently we obtain   
\begin{eqnarray}
	\partial_{r}F\,(r_{h},\theta)=\partial_{r}\Theta\,(r_{h},\theta)=0\, .\label{}
\end{eqnarray} 
The condition that the spacetime is asymptotically flat requires
\begin{eqnarray}
	 f\,(\infty,\theta)=m\,(\infty,\theta)=l\,(\infty,\theta)=1 \, . \label{}
\end{eqnarray}
The boundary conditions for $F$ and $\Theta $ at infinity remain the same 
as in flat spacetime      
\begin{eqnarray}
	F\,(\infty,\theta)=0\; ,\;\;\;\;\; \partial_{r}\Theta\,(\infty,\theta)=0 \, . \label{}
\end{eqnarray}
For the solution to be axially symmetric, we have 
\begin{eqnarray}
 	&&\partial_{\theta}f\,(r,0)
	=\partial_{\theta}m\,(r,0)
	=\partial_{\theta}l\,(r,0)=0 , \\
	&&\partial_{\theta}f\left(r,\frac{\pi}{2}\right)
	=\partial_{\theta}m\left(r,\frac{\pi}{2}\right)
	=\partial_{\theta}l\left(r,\frac{\pi}{2}\right)
	=0 . \label{}
\end{eqnarray}
Regularity on the axis and axisymmetry impose the boundary conditions on 
$F$ and $\Theta$ as  
\begin{eqnarray}
	&& \partial_{\theta}F\,(r,0)=\partial_{\theta}F\left(r,\frac{\pi}{2}\right)=0, \\
	&& \Theta\,(r,0)=0, \; \Theta\left(r,\frac{\pi}{2}\right)=\frac{\pi}{2} . \label{}
\end{eqnarray}
Under these boundary conditions, we shall solve the Einstein equations 
and the matter field equations numerically. 

\subsection{Numerical Results}

Let us introduce dimensionless coordinate and coupling constant   
\begin{eqnarray*}
	x=aF_{\pi}r \; , \;\;\;\;\;	\alpha = \pi G f_{\pi}^{2} \, .\label{}
\end{eqnarray*}
Then the free parameters are the horizon $x_{h}$ and the coupling constant $\alpha$ 
for this system.  
We shall take $\alpha = 0$ as decoupling of gravity from the matter, effectively $G=0$.

\begin{figure}
\hspace{2cm}
\includegraphics{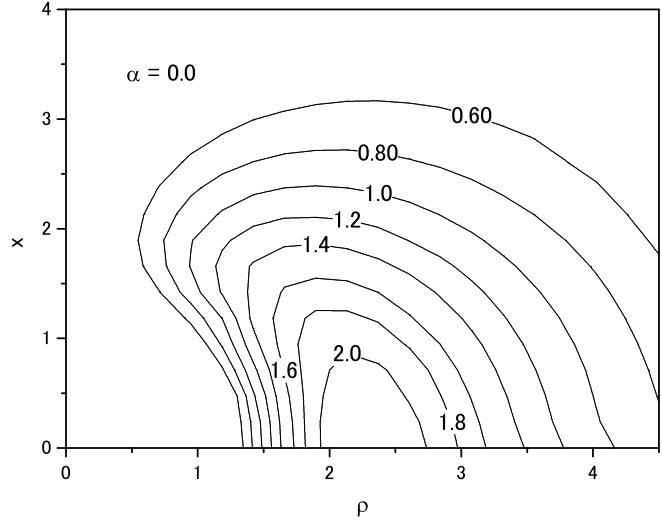}
\caption{\label{fig:ed0} The energy density $\epsilon$ 
in cylindrical coordinates $\rho$ and $z$ with $x_{h}=1.0, \alpha = 0.0$. } 
\end{figure}

\begin{figure}
\hspace{2cm}
\includegraphics{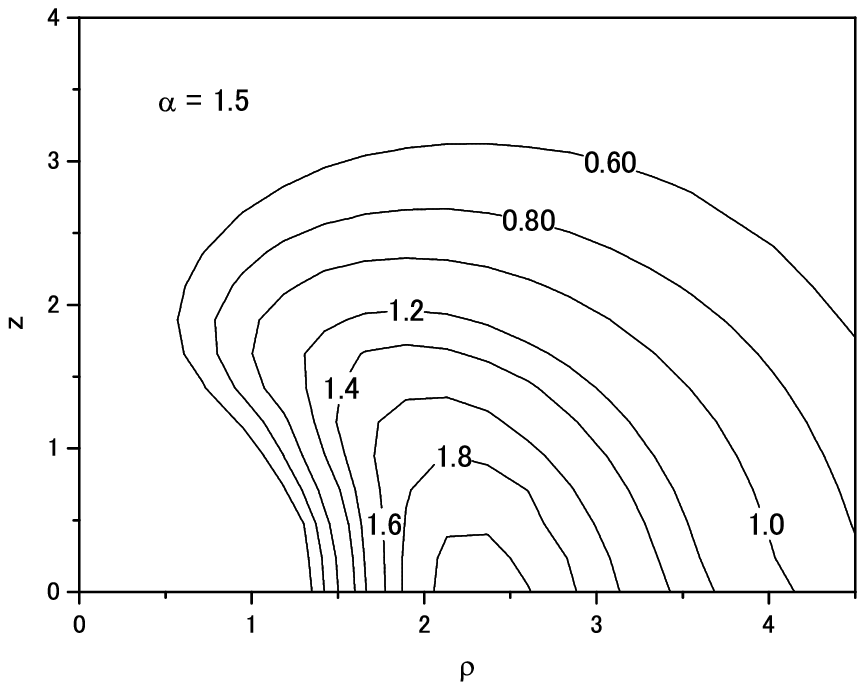}
\caption{\label{fig:ed15} The energy density $\epsilon$ 
in cylindrical coordinates $\rho$ and $z$ with $x_{h}=1.0, \alpha = 1.5$. } 
\end{figure}

\begin{figure}
\hspace{2cm}
\includegraphics{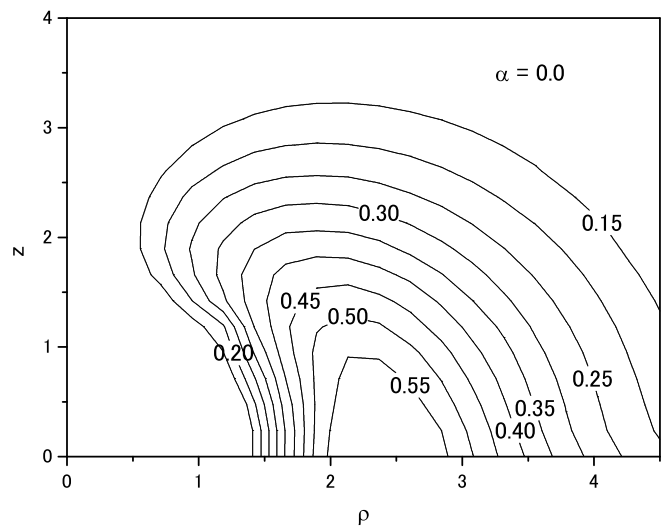}
\caption{\label{fig:b0} The baryon density $b$ with $x_{h}=1.0, \alpha =0.0$. }
\end{figure}

In Figs.~\ref{fig:ed0}, \ref{fig:ed15} are the energy densities of the black hole solutions 
with $\alpha = 0.0$, $1.5$ respectively. As $\alpha $ becomes larger, the energy density 
becomes smaller and sparse. This can be interpreted that the black hole absorbs more 
skyrmions  for a larger coupling constant. The shape is slightly distorted in the 
background of the black hole so that one can see the spherically symmetric horizon 
at the center of the skyrmion. Fig.~\ref{fig:b0} is the baryon density around 
the black hole. The dependence of the baryon density on the value of the coupling 
constant is small. It can be checked that the energy and baryon density 
vanish at $\rho =0 $. 
Inserting the metric functions as well as the profile functions 
expanded around the horizon instead, one can also see that the 
energy and baryon density vanish at the horizon.    

\begin{figure}
\hspace{2cm}
\includegraphics{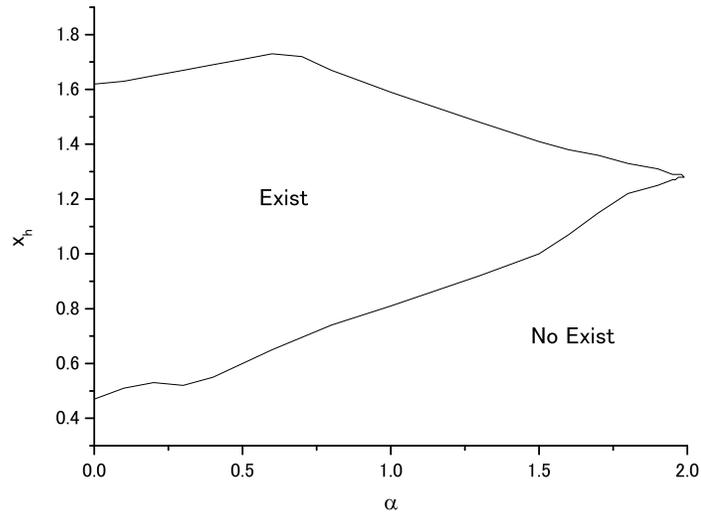}
\caption{\label{fig:domain} The domain of existence of the 
solution. For $\alpha \gtrsim 2.0$, there exists no non-trivial 
solution.}
\end{figure}

\begin{figure}
\hspace{2cm}
\includegraphics{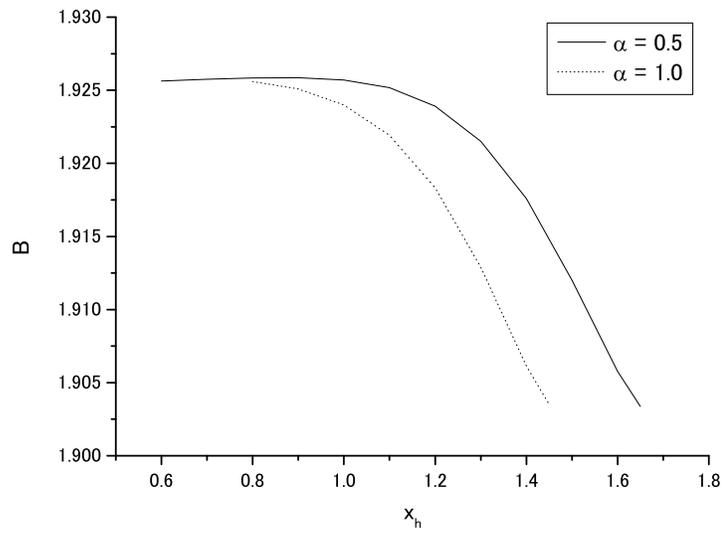}
\caption{\label{fig:baryon_number} The dependence of the baryon number 
on the size of the horizon.}
\end{figure}

The domain of existence of the black hole solution is shown in Fig.~\ref{fig:domain}. 
There exist minimum and maximum value of $x_{h}$ and $\alpha$ beyond which no 
black hole solutions exist. Therefore the regular skyrmion solutions can not be 
recovered from the black hole solutions by taking the limit of $x_{h} \rightarrow 0$ 
unlike the case of $B=1$~\cite{bizon92}.
In Fig.~\ref{fig:baryon_number} is the dependence of the baryon number on $x_{h}$. 
One can see that the baryon number decreases as the black hole grows in size.   
This figure confirms that the baryon number is no longer conserved due to the 
black hole absorbing the skyrmion. 


\pagebreak

\section{Conclusion and Discussion}

In this paper black hole solutions with Skyrme hair are reviewed. 
Sec.~3 is devoted to the study of the spherically symmetric black 
hole with $B=1$ Skyrme hair and its stability analysis. 
In Sec.~4 the gauged Einstein-Skyrme system is constructed  
and the monopole black hole with $B=1$ Skyrme hair is 
obtained. 
The extended Einstein-Skyrme system to $B=2$ axially symmetric 
configuration is studied in Sec.~5. 
The black hole solution with $B=2$ Skyrme hair is obtained 
and it is shown that the energy and baryon density exhibit 
a torus shape with the spherically symmetric horizon at 
the center.  

The common feature in those solutions is that they  
can support fractional baryonic charges outside the horizon, 
violating baryon number conservation. 
The study of the monopole black hole skyrmion shows that 
although global charge conservation such as baryon number 
is violated, gauge charge conservation such as electric or 
magnetic charge is still hold.  
It is remarkable that the black hole and monopole black hole 
solution with $B=1$ Skyrme hair turned out to be stable under 
linear perturbations even with the non-integer baryonic charge.        
Obviously it is important to study the stability of the $B=2$ solutions. 
We expect that the stability analysis may be performed by applying the 
catastrophe theory for black holes with non-linear hair~\cite{maeda9495}.  

For these microscopic black holes, however, we cannot ignore 
the quantum effects. In fact, they are stable only classically 
and will decay to the unstable solutions due to quantum transitions. 
Besides black holes of the size of a proton should have large fluxes 
of Hawking radiation~\cite{hawking75}. 
Hence situations in which baryon decay process might become more realistic and 
significant over the Hawking radiation occurs only when the black hole carries 
electric or/and magnetic charge with which the skyrmion interacts electromagnetically 
as well as gravitationally. Especially interesting is the extremal 
black hole which has a vanishing effective temperature, 
so the Hawking radiation may even vanish.    
The free skyrmion has a magnetic moment and, if it has
the correct orientation, it will be attracted to the monopole.
When the proton approaches the black hole monopole, the fields 
rearrange themselves into the energetically preferred configuration 
of skyrmion hair solutions described in Sec.~4. 
For the stable solutions, baryon decay can only 
takes place when extra particle fields are included.   
We have given a rigorous argument to study dynamical   
proton decay to a lepton by a monopole black hole. 
The bosonization technique is conveniently introduced 
and the system is reduced to two coupled time-dependent 
field equations. Solving these equations with appropriate boundary 
conditions at the horizon will be our future work.     

Recently we performed collective quantization of a $B=1$ gravitating 
skyrmion and calculated various nucleon observables 
under the gravitational effects in Ref.~\cite{shiiki04}. 
It will be interesting to extend the work to the black hole skyrmion.  
Also interesting but hard is to consider black holes with 
$B \ge 3$ Skyrme hair which has discrete symmetries. 
The rational map ansatz will be a powerful tool for obtaining 
multi-skyrmion black hole solutions.  
The inclusion of gauge fields may also be possible to study the interaction 
between a monopole black hole and multi-skyrmion. 
Extending the model to higher dimensions will be also an exciting problem. 
It is known that multi-extremal black hole solutions can be generalized 
to $p$-brane objects by coupling to an antisymmetric tensor $A_{p+1}$ with 
$p+1$ indices. In brane theory context, Skyrme fields are axions interacting 
with branes. The generalization of our model to brane theory will be worth 
studying in future.  

\begin{center}
  {\large {\bf Acknowledgement}}
\end{center}
We would like to thank I. G. Moss, E. Winstanley, T. Torii and K. Maeda 
for their collaboration and A. Flachi for useful discussions.

\end{document}